\documentclass[12pt,titlepage,twoside,fleqn]{article}

\usepackage[dvips]{pstcol}

\usepackage{graphicx}
\usepackage{amssymb}
\usepackage{amsmath}
\usepackage{cite}
\usepackage{here}
\usepackage{rotating}

\allowdisplaybreaks[1]

\def \beq{\begin{equation}}
\def \eeq{\end{equation}}
\def \bea{\begin{eqnarray}}
\def \eea{\end{eqnarray}}
\def \bit{\begin{itemize}}
\def \eit{\end{itemize}}
\def \nnl{\nonumber\\}             % nonumber and new line

\newcommand{\Lag}[1]{{\cal L}_{\rm #1}}          % Lagrangian density
\newcommand{\Li}[1]{{\rm Li}_2{\left(#1\right)}} % Dilogarithm function
\newcommand{\Cl}[1]{{\rm Cl}_2{\left(#1\right)}} % Clausen function

\def \dSt{\displaystyle}    % displaystyle
\def \tSt{\textstyle}       % textstyle

\def \Op{\mathcal{O}}         % operator
\def \Optilde{\tilde{\cal O}} % operator with tilde
\def \ord{O}                  % order of something
\def \gqcd{g_s}               % strong coupling constant
\def \alS{\alpha_s}           % strong coupling constant 
\def \alE{\alpha_e}           % electro-magnetic coupling constant
\def \sinWe{s_W}              % sin( theta_weinberg )

\def \lbl{{l\bar{l}}}                         % lepton leptonBar
\def \nbn{{\nu\bar\nu}}                       % neutrino neturinoBar
\def \bsgamma{b \to s \gamma}                 % b -> s gamma
\def \BXsgamma{\bar{B} \to X_s \gamma}        % B -> X_s gamma
\def \bsll{b \to s l^+l^-}                    % b -> s l+ l-
\def \BXsll{\bar{B} \to X_s l^+l^-}           % B -> X_s l+ l-
\def \BXsee{\bar{B} \to X_s e^+e^-}
\def \BXsmumu{\bar{B} \to X_s \mu^+\mu^-}
\def \BXclnu{\bar{B} \to X_c l \bar{\nu}}     % B -> X_c l nu
\def \BXulnu{\bar{B} \to X_u l \bar{\nu}}     % B -> X_u l nu (charmless)
\def \MSbar{\overline{\rm MS}}                % MS bar scheme
\def \Vckm{V_{\rm CKM}}                       % CKM - matrix
\def \Lqcd{\Lambda_{\rm QCD}}                 % Lambda_qcd
\def \Br{{Br}}                            % branching ratio
\def \AFBs{{\cal A}_{\rm{FB}}(\hat{s})}           % un-normalized forward-backward asymmetry
\def \barAFBs{\bar{{\cal A}}_{\text{FB}}(\hat{s})}  % normalized forward-backward asymmetry

\newcommand{\dgd}[1]{\gamma_{#1}}          % Dirac-Gamma-matrix sub index 
\newcommand{\dgu}[1]{\gamma^{#1}}          %                    raised sub index

\def \TeV{\mbox{TeV}}
\def \GeV{\mbox{GeV}}

\def \cIno{{\tilde{\chi}}}     % chargino
\def \gIno{{\tilde{g}}}        % gluino
\def \uSq{{\tilde{u}}}         % up-Squark
\def \dSq{{\tilde{d}}}         % down-Squark
\def \Sneu{{\tilde{\nu}}}      % Sneutrino

\newcommand{\XX}[4]{ [ {X_{#1}^{#2}}^{#4} ]_{#3}}  % chargino   - fermion - sfermion couplings
\newcommand{\MM}[3]{ {M^{#1}_{\tilde{#2}_{#3}}} }  % soft susy breaking masses % for scalars

\def \One{\leavevmode\hbox{\small1\kern-3.6pt\normalsize1}}  % unit matrix

\begin{document}

\begin{titlepage}

\begin{flushright}
{TUM-HEP-554/04  \\ 
 UCSD/PTH 04-13  \\
 hep-ph/0409293  \\ 
 September 2004}
\end{flushright}

\begin{center}
\vspace{2cm}

\setlength{\baselineskip}{0.3in}
{\LARGE \bf \boldmath
$\BXsll$  in the MSSM at NNLO }
\vspace{10mm} \\

\setlength{\baselineskip}{0.2in}
{\bf Christoph Bobeth$^{a,b}$, Andrzej~J.~Buras$^a$ 
     and Thorsten~Ewerth$^a$}\\
         
\vspace{3mm}
{\it $^a$
 Physik Department, Technische Universit\"at M\"unchen, D-85748 Garching, Germany}\\
\vspace{2mm}
{\it $^b$
 Physics Department, University of California at San Diego, La Jolla, CA 92093, USA}\\

\vspace{2cm}
{\large Abstract} \\ \vspace{1cm}

\begin{minipage}{14.5cm}
{\small 

We present the results of the calculation of QCD corrections to the matching
conditions for the Wilson coefficients of operators mediating the transition
$\bsll$ in the context of the MSSM. Within a scenario with decoupled heavy
gluino the calculated contributions together with those present already in the
literature allow for the first time a complete NNLO analysis of $\BXsll$.  We
study the impact of the QCD corrections and the reduction of renormalization
scale dependencies for the dilepton invariant mass distribution and the
forward-backward asymmetry in the inclusive decay $\BXsll$ restricting the
analysis to the ``low-$s$'' region and small values of $\tan\beta$.
The NNLO calculation allows to decrease the theoretical uncertainties
related to the renormalization scale dependence below the size of
supersymmetric effects in $\BXsll$ depending on their magnitude. While
it will be difficult to distinguish the MSSM expectations for the
branching ratio from the Standard Model ones, this can become possible
in the dilepton invariant mass distribution depending on the MSSM
parameters and $\hat{s}$. In this respect the position of the zero
of the forward-backward asymmetry $\hat{s}_0$ is even more promising.}
\end{minipage}
\end{center}
\vfill

\setlength{\baselineskip}{14pt}
\noindent \underline{\hspace{5cm}}\\
{\footnotesize *E-mail addresses: bobeth@su3.ucsd.edu, aburas@ph.tum.de, 
                                  tewerth@ph.tum.de}

\end{titlepage}

%%%%%%%%%%  end of title page  %%%%%%%%%%%%%%%%%%%%%%%%%%%%%%%%%%%%%%%%%%%%%%%%%

\setlength{\oddsidemargin}{-0.5cm}
\setlength{\evensidemargin}{0.5cm}
\setcounter{page}{1}
\setlength{\baselineskip}{18pt}

%%%%%%%%%%  introduction  %%%%%%%%%%%%%%%%%%%%%%%%%%%%%%%%%%%%%%%%%%%%%%%%%%%%%%
\renewcommand{\theequation}{\arabic{section}.\arabic{equation}}
\setcounter{equation}{0}

\section{\large Introduction}

The recent measurements of the branching ratio of the inclusive decay $\BXsll$
($l=e,\, \mu$) of the Belle Collaboration \cite{belle:incl:bsll} and the BaBar
Collaboration \cite{babar:incl:bsll} are expected to provide an important test
of the Standard Model (SM) and possible new physics effects at the electroweak
scale. Furthermore they allow for the extraction of informations complementary
to those from the radiative inclusive decay mode $\BXsgamma$ which is nowadays
well known both experimentally and theoretically in the SM and puts non-trivial
constraints on parameters of models beyond the SM.

In the discussion of the decay $\BXsll$ the major theoretical uncertainties
arise from the non-perturbative nature of intermediate $c\bar{c}$ states of the
decay chain $\bar{B}\to X_s J/\psi\to X_s l^+l^-$ and analogous higher
resonances. These decay channels interfere with the simple flavour changing
decay mechanism $\bar{B}\to X_s l^+l^-$ and the dilepton invariant mass
distribution can be only roughly estimated when the invariant mass of the lepton
pair $s\equiv q^2 = (p_{l^-} + p_{l^+})^2$ is not significantly away from
$M_{J/\psi}^2$ resulting in uncertainties larger than $\pm 20\%$
\cite{LW:prd53}. For this reason the charmonium decays are vetoed explicitly in
the experimental analysis \cite{belle:incl:bsll,babar:incl:bsll} by cuts on the
invariant dilepton mass around the masses of the $J/\psi$ and $\psi'$
resonances.

A rather precise determination of the dilepton invariant mass spectrum seems to
be possible once the values of $s$ are restricted to be below or above these
resonances.  Then the calculation can be performed using perturbative methods
whereas non-perturbative corrections can be addressed within the framework of
Heavy Quark Expansion (HQE).  However, contrary to the semileptonic decay
$\bar{B}\to X_{u,c} l\bar\nu_l$ and the radiative decay $\BXsgamma$ this method
is not applicable in the endpoint region of the spectrum as pointed out in
\cite{BI:npb525}. Here other approaches have to be used such as for example
Heavy Hadron Chiral Perturbation Theory (HH$\chi$PT) by summing over the
kinematically allowed exclusive channels to reliably estimate the magnitude of
the endpoint decay spectrum.

At the moment the low-$s$ region, accessible to $l=e$ and $\mu$, is
theoretically best understood. The non-perturbative corrections $(\Lqcd/m_b)^n$
to the dilepton invariant mass distribution are calculated up to the order $n=3$
\cite{FLS:prd49, AHHM:prd55, CRS:plb410, BI:npb525, BB:plb469} and turn out to
be small compared to the leading perturbative contribution -- however, still
involving poorly known matrix elements of the Heavy Quark Effective Theory
(HQET) for $n=3$. Furthermore the effects related to the tails of $c\bar{c}$
resonances in the low-$s$ region of the decay $\BXsll$ were estimated
model-independently by employing an expansion in inverse powers of the charm
quark mass in \cite{BIR:npb511} and the size of these $\Lqcd^2/m_c^2$
corrections was found to be similar to the size of $\Lqcd^2/m_b^2$
corrections. Because of the smallness of the non-perturbative corrections in the
low-$s$ region, the $\BXsll$ decay rate is precisely predictable up to about
$10\%$ uncertainty.

The calculations of the perturbative contribution
\cite{GSW:npb319,GDSN:plb223,CRV:plb258} up to the complete next-to-leading
order (NLO) in QCD \cite{M:npb393,BM:prd52} in the SM had not reached this
precision. In a series of recent papers the calculation was extended to the
next-to-next-to leading order (NNLO) in QCD being almost complete up to the
missing two-loop matrix element contributions of the four quark operators
$\Op_3$--$\Op_6$ which are expected to be small\footnote{ The analogous
corrections to $\BXsgamma$ are $1\%$ \cite{BCMU:npb631,AAH:hepph0401038}. }.
These calculations comprise
\begin{itemize}
\item
  corrections to the Wilson coefficients \cite{BMU:npb574},
\item
  the anomalous dimension matrices (ADM) of the renormalization group evolution
  of the Wilson coefficients
  \cite{CMM:plb400,GGH:npb673,GGH:in:prep,BGGH:jhep0404}\footnote{ The
  self-mixing of the four-quark operators $\Op_1$--$\Op_6$ is not published yet
  \cite{GGH:in:prep}, however the relevant result for $\BXsll$ can be found in
  \cite{BGGH:jhep0404}. }
\item
  the virtual and real corrections to the matrix elements of the relevant
  operators \cite{AAGW:plb507,AAGW:prd66,GHIY:npb685}.
\end{itemize}
Within the SM the inclusion of NNLO corrections reduces the branching ratios of
$\BXsee$ and $\BXsmumu$ by typically $12\%$ and $20\%$, respectively
\cite{ALGH:prd66}.  Furthermore uncertainties due to the dependence on the
renormalization scale of the top quark mass $\mu_t\sim M_W$ become reduced from
about $\pm16\%$ to $3\%$ \cite{BMU:npb574} and the inclusion of the NNLO matrix
element corrections decrease the low energy scale dependence $\mu_b\sim m_b$ from
$\pm13\%$ to a value about $\pm6.5\%$ \cite{AAGW:plb507,AAGW:prd66}. Furthermore
electroweak corrections were found to be a few percent \cite{BGGH:jhep0404}
removing the scale ambiguity of $\alE$ when going beyond LO.

Apart from the branching ratio and the dilepton invariant mass distribution, the
differential forward-backward asymmetry of leptons represents the third
interesting observable in the decay $\BXsll$. The leading contribution to the
forward-backward asymmetry arises in the SM at the NLO and thus the inclusion of
the NNLO corrections drastically reduces the renormalization scale dependence in
predictions of this observable. In particular it is very sensitive to new
physics effects and further, $\hat s_0=s_0/m_b^2$, the position at which the
forward-backward
asymmetry vanishes provides an important test of the SM \cite{B:prd52}. Within
the SM the inclusion of NNLO corrections in the evaluation of 
$\hat s_0$ leads to a
shift of $10\%$ to higher values accompanied by a reduction of the uncertainty
due to renormalization scale dependencies in the prediction from typically
$15\%$ to $5\%$ \cite{GHIY:npb648,ABGH:prd66,AAHP:mpla19}. The electroweak
corrections shift $\hat s_0$ by $+2\%$ \cite{BGGH:jhep0404}.

Clearly, in view of the improving experimental situation of the ongoing
$B$-physics dedicated experiments, such as the BaBar and Belle experiments,
hopefully the experimental uncertainties will decrease. Presently it is
desirable from the theoretical side to restrict future experimental analysis of
the dilepton invariant mass distribution to regions below and above the $c\bar
c$-resonances.

Besides testing the SM, once the experimental accuracy improves, the inclusive
decay $\BXsll$ will also allow to constraint models involving new physics
scenarios beyond the SM. The reliability of such constraints depend crucially on
theoretical uncertainties due to higher order corrections in the prediction of
observables as demonstrated by the SM analysis in the case of the importance of
NNLO QCD corrections. In the present work we report the results of a calculation
of QCD corrections to the matching conditions for the Wilson coefficients of
operators mediating the transition $\bsll$ in the context of the Minimal
Supersymmetric Standard Model (MSSM). We chose a scenario in which the
down-squark mass matrix decomposes into $2 \times 2$ matrices for each
generation and furthermore a heavy decoupled gluino within the MSSM parameter
space ensuring the completeness of the calculated QCD corrections.

The scenario of the MSSM that we study here has been introduced in
\cite{BEKU:prd66} as ``Scenario B'' (see also \cite{BCRS}). It is a
generalization of the ``Scenario A'' which was used in the context of the
calculation of the QCD corrections to $\bar B\to X_{s,d}\nu\bar\nu$, $\bar
B_{s,d}\to l^+l^-$, $K\to\pi\nu\bar\nu$ and $K_L\to\mu^+\mu^-$ in the MSSM
\cite{BBKU:npb630}\footnote{ The analytic results given in \cite{BBKU:npb630}
are applicable also in ``Scenario B'' considered here.}. In this paper QCD
corrections to the relevant $Z^0$ penguin diagrams, the box diagrams and the
neutral Higgs penguin diagrams have been calculated. While the results of these
calculations were ingredients of a NLO analysis of the decays considered there,
they contribute to $\BXsll$ first at the NNLO level and consequently enter our
present analysis. Actually as we concentrate on the region $\tan\beta \le 10$,
only $Z^0$ penguin diagrams and box diagrams calculated in \cite{BBKU:npb630}
are relevant here. The large $\tan\beta$ region where also neutral Higgs
penguins are relevant, will not be considered here.

Taking into account the results of \cite{BBKU:npb630}, the known results for the
Wilson coefficients of the magnetic penguins, the NNLO corrections to the matrix
elements of the relevant operators from
\cite{AAGW:plb507,AAGW:prd66,GHIY:npb685} and their three loop anomalous
dimensions calculated recently in
\cite{CMM:plb400,GGH:npb673,GGH:in:prep,BGGH:jhep0404}, the only missing
ingredients of a complete NNLO analysis of $\BXsll$ in the MSSM are the QCD
corrections to the Wilson coefficients of the four-quark operators
$\Op_1$--$\Op_6$ and the semileptonic operator $\Op_9$. These missing
ingredients are calculated here for the first time.

The main objectives of our paper are then as follows:
\begin{itemize}
\item
  the calculation of the matching conditions in question at $\ord(\alS)$
  which requires the evaluation of a large number of two-loop diagrams,
\item
  the calculation of the dilepton invariant mass distribution
  in $\BXsll$ and of the related
  forward-backward asymmetry in the MSSM at low $s$ and $\tan\beta$,
\item
  the investigation of the renormalization scale dependence of the observables
  in question and of the impact of the NNLO corrections on these observables in
  comparison with the NLO results,
\item
  the comparison of the NNLO results in the MSSM with those obtained in the SM,
\item
  the comparison of the size of MSSM corrections with the theoretical
  uncertainties in the SM.
\end{itemize}

The outline of this paper is as follows. In Section \ref{sect:MSSM} we briefly
review the elements of the MSSM relevant to the scenario with decoupled
gluinos. Section \ref{sect:wil:coeff} summarizes the low-energy effective
Lagrangian for the transition $\bsll$ and the corresponding Wilson coefficients
including $O(\alS)$ corrections in the context of the MSSM. Section
\ref{sect:form:obs} presents the formulae for the dilepton invariant mass
distribution and the forward-backward asymmetry of the leptons in the decay
$\BXsll$ including all NNLO corrections. The phenomenological implications for
both observables will be given in Section \ref{sect:num:ana}. We summarize and
conclude in Section 6.  Finally the appendices collect the analytical results of
the Wilson coefficients.

%%%%%%%%%%  MSSM scenario and gluino decoupling  %%%%%%%%%%%%%%%%%%%%%%%%%%%%%%
\renewcommand{\theequation}{\arabic{section}.\arabic{equation}}
\setcounter{equation}{0}

\section{\large \boldmath The considered Scenario of the MSSM\label{sect:MSSM}}

Let us start by specifying the scenario of the MSSM in which the analytical
calculation will be performed.\footnote{For the notation and conventions of
mixing matrices and couplings we will adopt see Section 2 of
\cite{BBKU:npb630}.}  First, we take the down-squark mass-squared matrix to be
flavour diagonal so that there are no neutralino contributions to
flavour-changing $b\to s$ transitions, and second, we assume the gluino with
mass $M_\gIno$ to be much heavier than all other sparticles. This first
assumption corresponds to ``Scenario B'' described in detail in
\cite{BEKU:prd66} which was also used in \cite{BCRS}. The second assumption
leads us to an ``effective MSSM'' with decoupled gluino at the scale
$\mu_\gIno\sim M_\gIno$ \cite{BMU:npb567}. Neglecting all the $1/M_{\tilde{g}}$
effects, the only modified couplings relevant for the NNLO corrections to
$\BXsll$ come from the ``chargino -- up-squark -- down-quark'' vertex,
\begin{align}
 X_i^{U_L} &= -g_2 \left[a_g\; V_{i1}^{\ast}\;\Gamma^{U_L}
    - a_Y\; V_{i2}^{\ast}\; \Gamma^{U_R}
    \frac{M_U}{\sqrt2 M_W \sin\beta}\right]V_{\rm CKM},\\[2mm]
 X_i^{U_R} &= g_2\; a_Y\;  U_{i2} \Gamma^{U_L}\;V_{\rm CKM}
  \frac{M_D}{\sqrt2 M_W \cos\beta},
\end{align}
where
\begin{align}\label{eff:gluino}
 a_g = 1-\frac{\alS(\mu_{\tilde{g}})}{4\pi}\left[ \frac{7}{3} 
       + 2\ln\left(\frac{\mu_{\tilde{g}}^2}{M_{\tilde{g}}^2}\right)\right], \quad
 a_Y = 1+\frac{\alS(\mu_{\tilde{g}})}{4\pi}\left[1+ 
        2\ln\left(\frac{\mu_{\tilde{g}}^2}{M_{\tilde{g}}^2}\right)\right].
\end{align}
It is at this scale $\mu_\gIno$ where these couplings as well as the up-squark
masses $m_\uSq$ and mixing matrices $\Gamma^{U_{L,R}}$ of the ``effective MSSM''
are determined in the matching with the full MSSM.
All of them are understood to be $\MSbar$ renormalized quantities in dimensional
regularization.
We refrain here from shifting the up-squark masses and mixing matrices of the
``effective MSSM'' into the on-shell scheme in order to avoid the appearance of
large logarithms ``$\ln(\mu_\gIno/m_{\uSq})$'', as can be seen by inspection
of (\ref{MS:OS:mass:shift}) and (\ref{MS:OS:GamU:shift}).
Then the next step is to integrate out successively all other particles with
masses much smaller than $M_\gIno$ and much larger than $m_t$ when going to smaller
scales using NLO renormalization group (RG) equations between all occurring matching
scales.
In our analysis, however, we integrate out all sparticles other than the gluino in
one step with the top quark, taking into account the LO RG running between $\mu_\gIno$
and $\mu_t$ for up-squark masses and their mixing matrices $\Gamma^{U}$.
Due to the quartic QCD-interaction of the scalar squarks the LO RG equations of masses
and mixing matrices are coupled and found to be
\begin{align}
  \mu \frac{d}{d\mu} m^2_{\uSq_a} & =
    \frac{\alS}{4\pi} \left[ - 8 m^2_{\uSq_a} + 
       \frac{8}{3} \sum_{b=1}^6 P^U_{ab} m^2_{\uSq_b} P^U_{ba} \right], \\
  \mu \frac{d}{d\mu} \Gamma^U_{ab} & =
    \frac{\alS}{4\pi} \frac{8}{3}
    \sum_{e=1}^6 \sum_{\substack{\scriptstyle c=1 \\ \scriptstyle c\neq a}}^6
    P^U_{ae} \frac{m^2_{\uSq_e}}{m^2_{\uSq_a} - m^2_{\uSq_c}} P^U_{ec} \Gamma^U_{cb},
\end{align}
with
\begin{align}
  P^U & = \Gamma^U \One^{\rm LR}_{6\times6} {\Gamma^U}^\dagger, &
 \One^{\rm LR}_{6\times6} \equiv {\rm diag}(1,1,1,-1,-1,-1).
\end{align}
The down-squark mixing matrix $\Gamma^D$ still retains its $2\times 2$ block
structure after scaling it down from $\mu_\gIno$ to $\mu_t$ using LO RG equations,
and thus neutralino contributions are absent in $b\to s+$(light particle) decays
in LO electroweak interactions at the scale $\mu_t$.

So far neither squark masses nor mixing matrices have been measured, and thus in
the numerical analysis we would like to vary the fundamental parameters of the
MSSM rather then the squark masses and mixing matrices of the ``effective
MSSM''.  Since the latter are determined from the former, when decoupling the
gluino in the $\MSbar$ scheme at the scale $\mu_\gIno$ such a RG evolution
becomes necessary when calculating the two-loop ``matrix- elements'' at the
scale $\mu_t$ when decoupling in a second step the heavy SM particles and the
remaining (apart from the gluino) sparticles.

%%%%%%%%%%  NLO matching conditions  %%%%%%%%%%%%%%%%%%%%%%%%%%%%%%%%%%%%%%%%%%%
\section{\large \boldmath The Two-Loop Matching Conditions
         \label{sect:wil:coeff}}
\setcounter{equation}{0}

The framework of effective theories applied to electroweak decays is a
convenient tool to resum QCD corrections to all orders using RG methods
\cite{AJBLH:BBL}. As explained in the previous section, the mass hierarchy of the SM
and the considered extension -- the ``effective MSSM'' -- allows for integrating
out the heavy degrees of freedom of masses $M_{\rm heavy} \geq M_W$. The effect
of the decoupled degrees of freedom will be contained in the Wilson coefficients
of the QCD and QED gauge invariant low-energy effective theory with five active
quark flavors.

The effective low-energy Lagrangian relevant to the inclusive decay $\BXsll$
resulting from the SM and the considered scenario of the MSSM has the following
form
\begin{align}\label{effLag:bsll}
\Lag{eff} & = 
     \Lag{QCD\times QED}(u,d,s,c,b,e,\mu,\tau) 
\nnl[3mm]
  & + \frac{4 G_{\rm F}}{\sqrt2} 
     \left[  V_{us}^{\ast}V_{ub}^{}\, (C_1^c \Op_1^u + C_2^c \Op_2^u)
           + V_{cs}^{\ast}V_{cb}^{}\, (C_1^c \Op_1^c + C_2^c \Op_2^c)
     \right] 
\nnl[3mm]
  & + \frac{4 G_{\rm F}}{\sqrt2} \sum_{i \in A} 
     \left[(  V_{us}^{\ast}V_{ub}^{} + V_{cs}^{\ast}V_{cb}^{} )\, C_i^c 
            + V_{ts}^{\ast}V_{tb}^{}\, C_i^t\right] \Op_i,
\end{align}
with $A=\{3 \dots 10, \; EOM \, vanishing, \; evanescent\}$ numbering the
relevant operators $\Op^Q_i$ and the corresponding Wilson coefficients
$C^Q_i$. Here $G_F$ is the Fermi constant and furthermore we refrain from using
unitarity of the CKM matrix.  The first term in (\ref{effLag:bsll}) consists of
kinetic terms of the light particles -- the leptons and the five light quark
flavours -- as well as their QCD and QED interactions while the remaining terms
consist of $\Delta B= -\Delta S = 1$ gauge-invariant local operators\footnote{
The operators conserve flavours other than $B$ and $S$.}  up to dimension 6 built
out of those light fields\footnote{The $s$-quark mass is neglected here,
i.e. it is assumed to be negligibly small when compared to $m_b$.}.
The operators $\Op^Q_i$
entering the effective Lagrangian can be divided into three classes.

The {\em physical} operators are
\begin{align} 
  \Op_1^Q & = (\bar{s}\,\dgd{\mu} P_L {\bf T}^a\, Q)
    (\Bar{Q}\,\dgu{\mu} P_L {\bf T}^a\, b), &
  \Op_2^Q & = (\bar{s}\, \dgd{\mu} P_L\, Q)
    (\Bar{Q}\, \dgu{\mu} P_L\, b), \nnl[4mm]
%%%  
  \Op_3 & = (\bar{s}\, \dgd{\mu} P_L\, b) \,
     \sum_q (\bar{q}\, \dgu{\mu}\, q), & 
  \Op_5 & = (\bar{s}\, \dgd{\mu_1}\dgd{\mu_2}\dgd{\mu_3} P_L\, b) 
      \sum_q (\bar{q}\, \dgu{\mu_1}\dgu{\mu_2}\dgu{\mu_3} \, q), \nnl
%%%
  \Op_4 & = (\bar{s}\, \dgd{\mu} P_L {\bf T}^a\, b) \,
      \sum_q (\bar{q}\, \dgu{\mu} {\bf T}^a\, q), &
  \Op_6 & = (\bar{s}\, \dgd{\mu_1}\dgd{\mu_2}\dgd{\mu_3} P_L {\bf T}^a\, b) 
      \sum_q (\bar{q}\, \dgu{\mu_1}\dgu{\mu_2}\dgu{\mu_3} {\bf T}^a\, q), \nnl 
%%%
  \Op_7 & = \frac{e}{\gqcd^2} m_b \;
      (\bar{s}\, \sigma^{\mu\nu} P_R\, b) \; F_{\mu\nu}, &
  \Op_8 & = \frac{1}{\gqcd}   m_b \;
      (\bar{s}\, \sigma^{\mu\nu} P_R {\bf T}^a\, b)\; G^a_{\mu\nu}, \nnl[1mm]
%%%
  \Op_9 & = \frac{e^2}{\gqcd^2} (\bar{s}\, \dgd{\mu} P_L \,b)
      (\bar{l}\, \dgu{\mu} \, l), &
  \Op_{10} & = \frac{e^2}{\gqcd^2} (\bar{s}\, \dgd{\mu} P_L \,b)
      (\bar{l}\, \dgu{\mu} \dgd{5} \, l),
\label{phys:op:basis}
\end{align}
where $P_{L,R}=\frac{1}{2}(1\mp\gamma_5)$ are the left- and right-handed chirality
projectors, respectively. They
consist of the current-current operators $\Op^Q_{1,2}$ ($Q=\{u,c\}$), the QCD
penguin operators $\Op_{3,\ldots,6}$ ($q=\{u,d,s,c,b\}$), the electro- and
chromo-magnetic moment type operators $\Op_{7,8}$ and finally the semileptonic
operators $\Op_{9,10}$. It should be noted that the above basis of physical
operators results from the SM, however in extensions of the SM other physical
operators could become relevant, too. In the MSSM scenario chosen here this is
not the case for low values of $\tan\beta$ and the SM operator basis suffices.

In addition to the physical operators several non-physical operators have to be
included in the matching procedure of the full and effective theories. The
so-called {\em EOM vanishing} operators that vanish by the QCD$\times$QED
equation of motion (EOM) of the effective theory up to a total derivative can be
found in Section 5 of \cite{BMU:npb574}. They appear in intermediate steps of the
off-shell calculation of the processes $\bsgamma$ and $b\to s g$ and contribute
to the final results of Wilson coefficients of physical operators when going
beyond leading order matching.

The second group of non-physical operators which have to be considered in the
matching procedure are {\em evanescent} operators. Evanescent operators vanish
algebraically in four dimensions, however in $D\neq 4$ dimensions they are
indispensable and contribute to Wilson coefficients of physical operators. We
use the same convention for the evanescent operators as introduced in the
evaluation of the anomalous dimensions relevant to $\bsgamma$, $b\to sg$ and
$\bsll$ of \cite{CMM:plb400,GGH:npb673}.

The specific structure of the operators $\Op_i$ is determined from the
requirement that the effective theory reproduces the SM $\Delta B=-\Delta S=1$
off-shell amplitudes of $b\to s+$(light particles) at the leading order in
electroweak gauge couplings and up to $O[$(external momenta and light
masses)$^2$/$M_{\rm heavy}^2 ]$), but to all orders in strong interactions. The
same applies to the extensions of the SM. For a detailed description of the
two-loop matching of photonic $\Delta B = 1$ penguins ($\bsgamma$) in the SM we
refer the interested reader to Section 5 of \cite{BMU:npb574}. The matching
calculation of the supersymmetric contributions is performed analogously. Here
in addition helpful details can be found in Section 4 of \cite{BBKU:npb630}.

The Wilson coefficients at the matching
scale $\mu_t$ can be perturbatively expanded in $\alS(\mu_t)$ as follows
\beq
  \label{wc:exp}
  C_i^Q = C_i^{Q(0)} + \frac{\alS}{4\pi} C_i^{Q(1)} + \frac{\alS^2}{(4\pi)^2}
          C_i^{Q(2)} + \ldots\; , \qquad Q=\{c,t\}.  
\eeq 
Contributions to order $\alS^n$ to each Wilson coefficient originate
from $n$-loop diagrams which follows from the particular convention of
powers of the QCD gauge coupling $\gqcd = \sqrt{4 \pi \alS}$ in the
normalization of the operators $\Op_{7,\ldots,10}$ in
(\ref{phys:op:basis}).

\begin{figure}
\begin{center}
  \scalebox{0.75}{\includegraphics{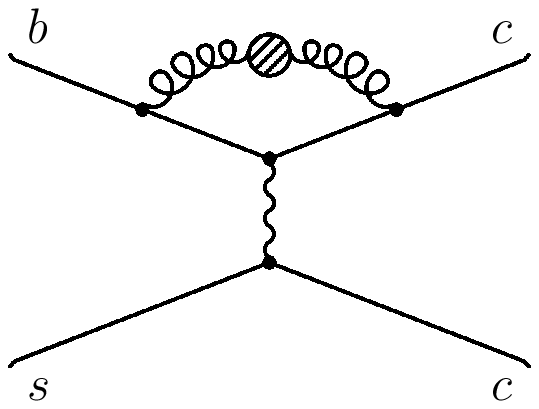}} \hspace{4mm}
  \scalebox{0.75}{\includegraphics{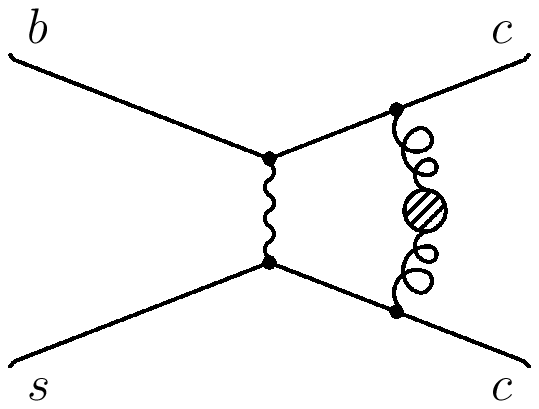}} \hspace{4mm}
  \scalebox{0.75}{\includegraphics{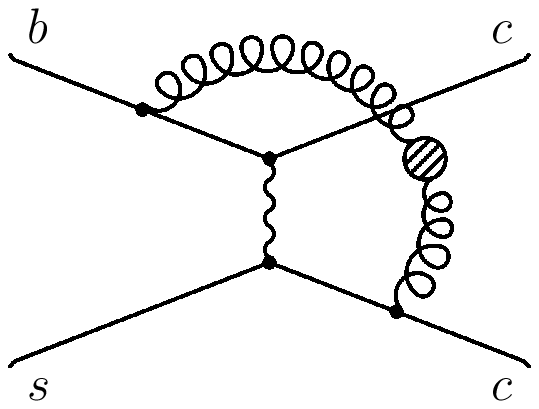}}
 \caption{\sf Two-loop contributions to the function $[T_1]^1_{\tilde{q}}$.
   The wiggly line denotes the $W$ boson.
   Shaded blobs stand for self-energy insertions with up- and down-squarks
   in the loop.
   Possible left-right and up-down reflected diagrams are not shown.}
 \label{fig:bscc:2l}
\end{center}
\end{figure}

The result of the matching computation of the Wilson coefficients of the
physical operators $\Op^Q_{1,\ldots,10}$ can be summarized as follows. At the
tree-level the only nonzero Wilson coefficient is $C_2^{c(0)} = -1$.
At the one- and two-loop level, the only matching condition in the
``charm-sector'' which gets contributions from virtual exchange of sparticles
(see Fig.~\ref{fig:bscc:2l}) is
\bea\label{charm:wc}
  C_1^{c(2)} & = [T_1]^1_t + [T_1]^1_{\tilde{q}}
  -\dSt\frac{7987}{72} - \dSt\frac{17}{3} \pi^2 - 
  \dSt\frac{475}{6} \ln\dSt\frac{\mu_W^2}{M_W^2} - 17
  \left[\ln\dSt\frac{\mu_W^2}{M_W^2}\right]^2, 
\eea
with $\mu_W$ being the renormalization scale in the ``charm-sector''. In the
notation of \cite{BMU:npb574}, $[T_1]^1_t = T(x)$ with $x = m_t^2/M_W^2$ is the
SM top quark contribution. Due to the chosen renormalization prescription the first
diagram given in Fig.~\ref{fig:bscc:2l} is completely ``renormalized away''.
Thus $C_2^{c(2)}$ is not affected by virtual sparticle exchange.  The last two
diagrams contribute to $[T_1]^1_{\tilde{q}}$ for which we obtain\footnote{ Here we
assumed $m_{\tilde{q}_a} > M_W/2$ which is clearly fulfilled.}
\begin{align}\hspace{-8mm}
 {[T_1]}^1_{\tilde{q}} &= \sum_{a=1}^6 \sum_{q=u,d} \left\{ 
    2 (4 x_{\tilde{q}_a} - 1)^\frac{3}{2} \;
    \Cl{2 \arcsin \tSt \frac{1}{2\sqrt{x_{\tilde{q}_a}}}} 
    - 8\left(x_{\tilde{q}_a} - \frac{1}{3}\right) \ln  x_{\tilde{q}_a}
    - 16 x_{\tilde{q}_a} \right\} + \frac{208}{3},
\end{align}
where $x_{\tilde{q}_a}=m_{\tilde{q}_a}^2/M_W^2$, and the definition of the Clausen
function $\Cl{x}$ can be found in \ref{app:wil:coeff}. As far as the remaining matching
conditions in the ``charm-sector'' and the
function $[T_1]^1_t$ are concerned we refer the reader to \cite{BMU:npb574}.

\begin{figure}%[H]
\begin{center}
  \scalebox{0.8}{\includegraphics{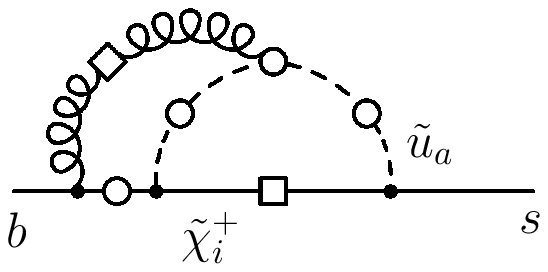}} \hspace{4mm}
  \scalebox{0.8}{\includegraphics{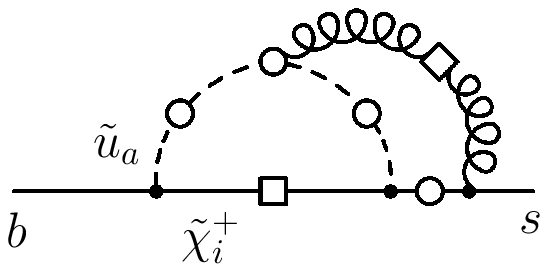}} \hspace{4mm}
  \scalebox{0.8}{\includegraphics{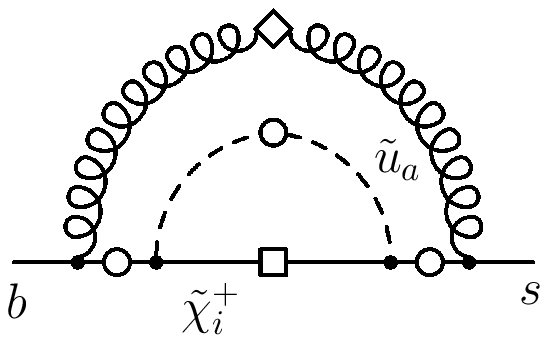}} \\
  \scalebox{0.8}{\includegraphics{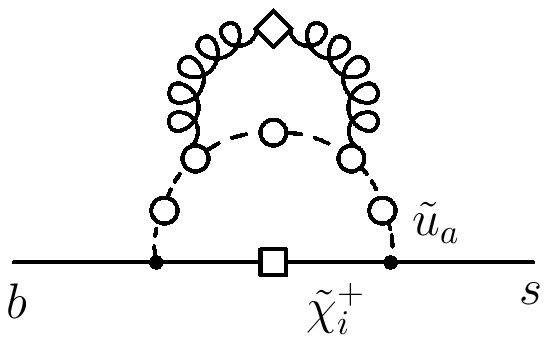}} \hspace{4mm}
  \scalebox{0.8}{\includegraphics{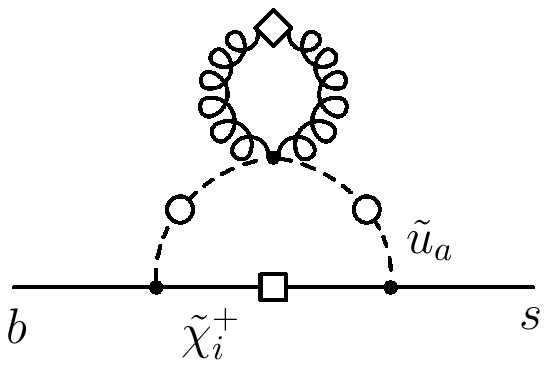}} \hspace{4mm}
  \scalebox{0.8}{\includegraphics{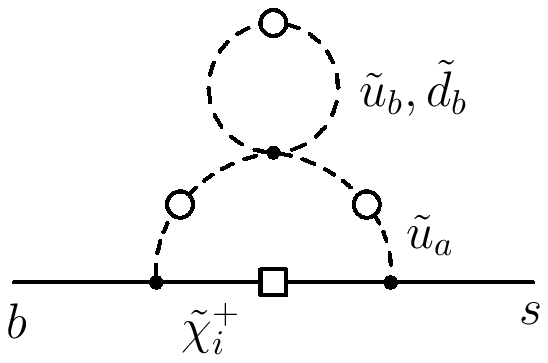}}
 \caption{\sf Two-loop contributions to the functions $[D_9]^1_i,\,[E_4]^1_i$
   with $i=\cIno,4$.
   The circle indicates the positions where photons or gluons can be emitted,
   whereas the square (diamond) indicates the positions where only photons (gluons)
   can be emitted.}
 \label{fig:bsg:2l}
\end{center}
\end{figure}

\begin{figure}%[H]
\begin{center}
  \scalebox{0.8}{\includegraphics{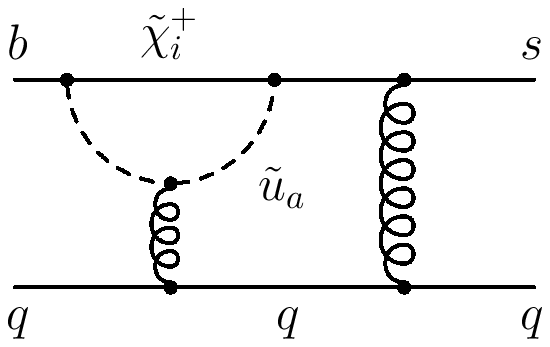}} \hspace{4mm}
  \scalebox{0.8}{\includegraphics{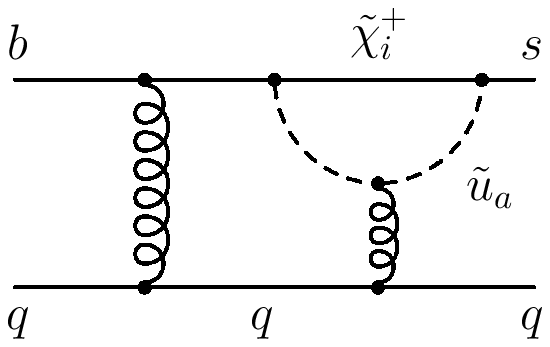}} \\
  \scalebox{0.8}{\includegraphics{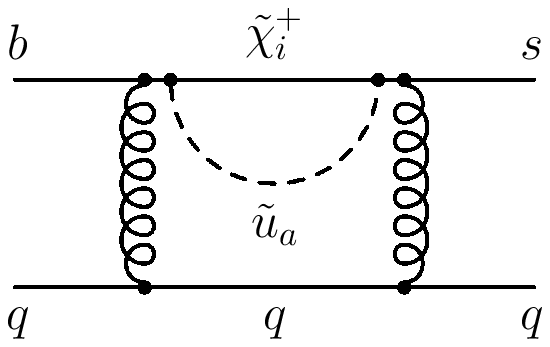}} \hspace{4mm}
  \scalebox{0.8}{\includegraphics{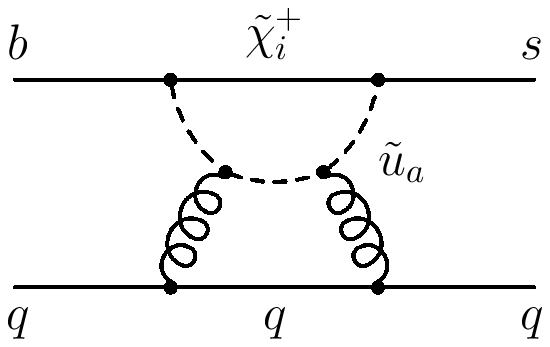}} \hspace{4mm}
  \scalebox{0.8}{\includegraphics{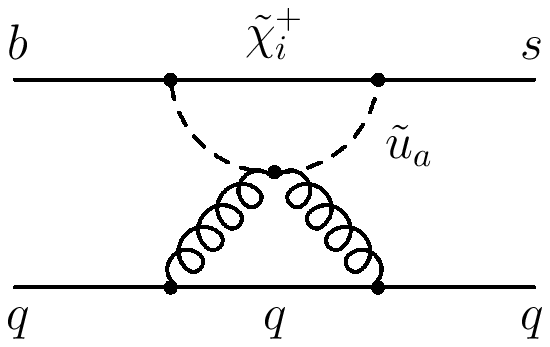}}
 \caption{\sf Two-loop contributions to the function $[G_3]^1_i$ with $i =\cIno,4$.
   Diagrams with crossed gluon lines are not shown.}
 \label{fig:bsqq:2l}
\end{center}
\end{figure}

The one-loop and two-loop matching conditions in the ``top-sector'' are
\begin{align}\label{top:wc}
  C_3^{t(1)} & = 0, \hspace{4cm}C_3^{t(2)} = [G_3]^1,  
\nnl[1mm]
  C_4^{t(n)} & = [E_4]^{(n-1)}, 
\nnl[1mm]
  C_5^{t(1)} & = 0, \hspace{4cm}C_5^{t(2)} =
    -\frac{1}{10} [G_3]^1 + \frac{2}{15} [E_4]^0, 
\nnl[1mm]
  C_6^{t(1)} & = 0, \hspace{4cm}C_6^{t(2)} =
    -\frac{3}{16} [G_3]^1 + \frac{1}{4} [E_4]^0, 
\nnl[1mm]
  C_7^{t(n)} & = -\frac{1}{2} [A_7]^{(n-1)}, 
\nnl[1mm]
  C_8^{t(n)} & = -\frac{1}{2} [F_8]^{(n-1)}, 
\nnl[1mm]
  C_9^{t(n)} & = \frac{1 - 4 \sinWe^2}{\sinWe^2} [C_9^\lbl]^{(n-1)} -
     \frac{1}{\sinWe^2} [B_9^\lbl]^{(n-1)} - [D_9]^{(n-1)}, &&
\nnl[1mm]
  C_{10}^{t(n)} &= \frac{1}{\sinWe^2} ([B_{10}^\lbl]^{(n-1)}
     - [C_9^\lbl]^{(n-1)}).
\end{align} 
The various functions $[X]^n$ in (\ref{top:wc})
indicate their origin when matching the $b\to s+$(light particles) Greens
functions of the full and effective theory
\bit
\item $[A]$: on-shell part of 1PI $\bsgamma$ (see Fig.~\ref{fig:bsg:2l}),
\item $[B^\lbl]$: $\bsll$ mediated by box-diagrams,
\item $[C^\lbl]$: $\bsll$ mediated by $Z^0$ penguin diagrams,
\item $[D]$: off-shell part of 1PI $\bsgamma$, contributing to $\bsll$
      (see Fig.~\ref{fig:bsg:2l}),
\item $[E]$: off-shell part of 1PI $b\to s g$, contributing to $b\to sq\bar{q}$
      (see Fig.~\ref{fig:bsg:2l}),
\item $[F]$: on-shell part of 1PI $b\to s g$ (see Fig.~\ref{fig:bsg:2l}),
\item $[G]$: 1PI two-loop diagrams $b\to sq\bar{q}$ (see Fig.~\ref{fig:bsqq:2l}).
\eit 
The index $n$ corresponds to the number of loops in the diagrams which can be
classified into tree-level ($n=0$), NLO ($n=1$) and NNLO ($n=2$) contributions, 
see also the comment below (\ref{wc:exp}). Furthermore each function $[X]^n$ receives
contributions from different virtual particle exchange

\beq
  \label{Xfunctions}
  [X]^n = \sum_{i = \{W, H, \tilde\chi, 4\}} [X]^n_i.
\eeq
The index $i$ corresponds to
\bit
\item $i=W$: ``top quark -- $W$ boson'' loops (SM),
\item $i=H$: ``top quark -- charged Higgs boson'' loops,
\item $i=\cIno$: ``chargino -- up-squark'' loops,
\eit
receiving virtual gluon corrections at NNLO and further
\bit 
\item $i=4$: ``chargino -- up-squark'' loops including the 
      quartic squark vertex corrections proportional to $\gqcd$. 
      These diagrams contribute only at NNLO.
\eit
Discarding the contributions $\{H, \cIno, 4\}$ in the sum of
(\ref{Xfunctions}) one recovers the SM results, whereas
discarding only $\{\cIno, 4\}$ one obtains the results for Two Higgs
Doublet Models (2HDM) of type II provided $\tan\beta$ is small.

Explicit expressions for the various functions $[X]^n$ can be found in
\ref{app:wil:coeff}. We stress that all parameters appearing there are $\MSbar$
renormalized. To obtain the Wilson coefficients in terms of on-shell masses
and mixing matrices for squarks, the following steps should
be performed:
\begin{enumerate}
\item 
  Remove the contributions due to strong quartic squark couplings, i.e.~the
  contributions with the index $i=4$ in the functions $[X]^n$.
\item 
  Make the following shift of the up-squark mass in the contributions with the
  index $i=\cIno$: 
  \beq {m}^2_{{\tilde u}_a}(\mu) = 
       (m^{\rm pole}_{{\tilde u}_a})^2 \Bigg\{1- 
        \frac{\alpha_s(m^{\rm pole}_{{\tilde u}_a})}{4\pi}
        \frac{4}{3}\; \Bigg[7+ 3\ln \Bigg(
        \frac{\mu}{m^{\rm pole}_{{\tilde u}_a}}\Bigg)^2\Bigg]\Bigg\}, 
  \eeq 
  Observe that this shift involves only the gluonic corrections, since the
  contributions due to strong quartic squark couplings have already been
  considered in step 1.
\end{enumerate}
The above two steps are a direct consequence of the application of the full
scheme shift from the $\MSbar$ to the on-shell scheme given in
(\ref{MS:OS:mass:shift}) and (\ref{MS:OS:GamU:shift}).

However, using the Wilson coefficients in terms of on-shell quantities one needs
of course on-shell input parameters. In our approach (see Section \ref{sect:MSSM}) we
have $\MSbar$ quantities at the scale $\mu_t$, and shifting them to their on-shell
values with the help of (\ref{MS:OS:mass:shift}) and (\ref{MS:OS:GamU:shift}) only
reproduced our numerical results in the $\MSbar$ scheme if all squark masses are close
in size. More properly one should integrate out squarks stepwise if their mass splittings
are large, and then shift to the on-shell scheme at the appropriate scale for each squark.
We chose to integrate out all squarks at one scale, and hence we refrain from working
in the on-shell scheme in our numerical analysis.

To summarize:
\bit
\item
  The contributions $[A_7]^1_i,\,[F_8]^1_i,\,[C_9]^1_i,\,[B_{10}]^1_i$ with $i =
  W, H, \cIno, 4$ and $[D_9]^1_W,\,[B_9]^1_W$, $[E_4]^1_W,\,[G_3]^1_W,\,[T_1]^1_t$
  have been calculated previously with the
  list of references given in \ref{app:wil:coeff}.
\item
  The contributions $[D_9]^1_i,\,[B_9]^1_i,\,[E_4]^1_i,\,[G_3]^1_i$
  with $i = H,\cIno,4$ and $[T_1]^1_{\tilde{q}}$ have been calculated here
  for the first time with the
  expressions listed in \ref{app:wil:coeff}. The contributions with $i=H$ have
  been calculated already in \cite{B:PhD} and very recently the same result
  has been obtained in \cite{SGST:hepph0407323}.
\eit

%%%%%%%%%%  Observables  %%%%%%%%%%%%%%%%%%%%%%%%%%%%%%%%%%%%%%%%%%%%
\renewcommand{\theequation}{\arabic{section}.\arabic{equation}}
\setcounter{equation}{0}

\section{\large Differential Decay Distributions 
         \label{sect:form:obs} }

In this Section we provide the formulae of some differential decay
distributions of the decay $\BXsll$. These are the dilepton invariant
mass spectrum and the differential forward-backward asymmetry with
respect to the dilepton invariant mass $s$ of the lepton
pair. They are given in terms of the Wilson coefficients at the
low-energy scale $\mu_b \sim m_b$ which are obtained by solving the
RG equation \cite{CMM:plb400,BMU:npb574,GGH:npb673} and the matrix elements of
the operators of the low-energy effective theory. At the scale $\mu_b$
usually the rescaled operators $\Optilde_i = \alS/(4\,\pi)\; \Op_i\;
(i = \{7,\,8,\,9,\,10\})$ are used. The corresponding Wilson coefficients
are $\tilde{C}_i(\mu_b) = (4\,\pi)/\alS\; C_i(\mu_b)$. 

The method of the HQE is applicable to the inclusive decay $\BXsll$
predicting the leading contribution to be the matrix elements of the
quark-level transition $\bsll$ whereas non-perturbative corrections of
the type $(\Lqcd/m_b)^n$ can be taken systematically into
account. However, this method is not applicable over the whole
kinematical range of $s$ and in this work we will restrict the
analysis to the so-called low-$s$ region \cite{BMU:npb574} below the
$c\bar c$-resonances.

The matrix elements of the four-quark operators $\Op_i\; (i=1,\dots6)$
to the process $\bsll$ are proportional to the tree-level matrix
elements of $\Optilde_7$ and $\Optilde_9$. It has become customary to
take them into account by the introduction of the effective Wilson
coefficient $\widetilde C_7^{{\rm eff}}$ and $\widetilde C_9^{{\rm
eff}}$. The exact expressions for these effective coefficients
relevant for the NLO analysis can be found in
\cite{M:npb393,BM:prd52,BMU:npb574}, whereas the NNLO corrections are
given in \cite{AAGW:plb507} for low values of $s$. This involves an
expansion in the ratios $m_c/m_b,\; \sqrt{s}/m_b$ and $\sqrt{s}/2
m_c$.  The calculation valid for all dilepton invariant masses $s$ can
be found in \cite{GHIY:npb685}.

The dilepton invariant mass spectrum with respect to the normalized
dilepton invariant mass $\hat{s} \equiv s/m^2_b$ reads
\begin{align}
 &\frac{d\Gamma(\bsll)}{d\hat{s}} =
  \left(\frac{\alE}{4\,\pi}\right)^2
  \frac{G_F^2\,(m_b^{\rm pole})^5\left|V_{ts}^*V_{tb}^{}\right|^2} 
  {48\,\pi^3}(1-\hat{s})^2 \nnl[1mm]
 &\,\times\left\{\left(1+2\,\hat{s}\right)\left(
  \left|\widetilde C_9^{{\rm eff}}\right|^2 +
  \left|\widetilde C_{10}^{{\rm eff}}\right|^2 \right)\,
  [1+\frac{2\,\alpha_s}{\pi}\,\omega_{99}(\hat{s})]
  + 4\left(1+\frac{2}{\hat{s}}\right)
  \left|\widetilde C_7^{{\rm eff}}\right|^2
  [1+\frac{2\,\alpha_s}{\pi}\,\omega_{77}(\hat{s})]\right.\nnl[1mm]
 &\hspace{1cm}\left.
  +12\,\mbox{Re}\left (\widetilde C_7^{{\rm eff}}
  \widetilde C_9^{{\rm eff}*}\right)
  [1+\frac{2\,\alpha_s}{\pi}\,\omega_{79}(\hat{s})]\right\}
  +\frac{d\Gamma^{\rm{Brems,A}}}{d\hat{s}}
  + \frac{d\Gamma^{\rm{Brems,B}}}{d\hat{s}}.
  \label{bsll:diff:Gamma}
\end{align}
The functions $\omega_{ij}(\hat{s})$ summarize the virtual and real QCD
corrections to the matrix elements of the operators $\Optilde_i \; (i =
\{7,\,9,\,10\})$ \cite{AAGW:plb507,GHIY:npb648}, whereas the terms
$d\Gamma^{\rm{Brems,A}}/d\hat{s}$ and
$d\Gamma^{\rm{Brems,B}}/d\hat{s}$ result from infrared-finite real
corrections \cite{AAGW:prd66}. In the numerical analysis we follow
\cite{BGGH:jhep0404} concerning the QCD corrections.  However, we will
not include the higher order QED corrections discussed there, but
rather use $\alE = 1/133$ which yields results close to the once
obtained including them, as was found in \cite{BGGH:jhep0404}.

To obtain the hadronic differential decay rate $d\Gamma(\BXsll)/d\hat{s}$
within HQE, $(\Lqcd/$ $m_b)^n$ corrections
have to be added to the partonic differential decay rate
$d\Gamma(\bsll)/d\hat{s}$ of (\ref{bsll:diff:Gamma})
\cite{FLS:prd49,AHHM:prd55,CRS:plb410,BI:npb525,BB:plb469}. These
corrections were calculated up to the order $n=3$
\cite{BB:plb469}. In the numerical analysis we will only include the corrections
$n=2$ as the corrections $n=3$ involve poorly known hadronic matrix elements. We
also include the $(\Lqcd/m_c)^2$ corrections of \cite{BIR:npb511}.

The partially integrated branching ratio $\Br(\BXsll)$ of the low-$s$ region is
\beq\label{br:BXsll}
  \Br(\BXsll) = \frac{\Br(\bar{B}\to X)}{\Gamma(\bar{B}\to X)} 
    \int_{q^2_{\rm min}}^{q^2_{\rm max}} 
    \frac{d\Gamma(\BXsll)}{d\hat{s}} \frac{d(q^2)}{m_b^2}
\eeq
with the boundaries chosen to be $q^2_{\rm min} = 1 \GeV^2$ and $q^2_{\rm max} =
6 \GeV^2$. A very recent experimental result of Belle for this quantity can be
found in the second paper of \cite{belle:incl:bsll} $\Br(\BXsll) = (1.49 \pm
0.50^{+0.38}_{-0.28})\times 10^{-6}$ which is in agreement with the BaBar measurements
in the second paper of \cite{babar:incl:bsll} both having comparable errors.

Commonly the semileptonic decay $\BXclnu$ is used as normalization because the
factor $(m_b^{\rm pole})^5$ -- the origin of large uncertainties -- cancels in
the ratio. An alternative was proposed in \cite{GM:npb611} using the charmless
semileptonic decays $\BXulnu$ and $\BXclnu$ in the calculation of the inclusive
decay $\BXsgamma$ reducing the uncertainties due to the charm quark mass $m_c$
present in $\Gamma(\BXclnu)$. The application of this method to $\BXsll$ can be
found in \cite{BGGH:jhep0404,CS:epjc33} and will be used in the numerical 
analysis. 

The so-called un-normalized forward-backward asymmetry is defined as
\beq
 \label{AFB:unnorm}
 \AFBs = \frac{\Br(\bar{B}\to X)}{\Gamma(\bar{B}\to X)}
 \int_{-1}^1\frac{d^2\Gamma(\BXsll)}{d\hat{s}\,dz}\,{\rm{sgn}}(z)\,dz.
\eeq
Again the normalization is commonly chosen to be the semileptonic decay
$\BXclnu$, however also the alternative of the combination of the decays
$\BXulnu$ and $\BXclnu$ \cite{BGGH:jhep0404} to reduce the uncertainties due to
$m_c$ can be applied. The so-called normalized forward-backward asymmetry is
given by the ratio
\beq
 \label{AFB:norm}
 \barAFBs = 
 \int_{-1}^1 \frac{d^2\Gamma(\BXsll)}{d\hat{s}\,dz}\,{\rm{sgn}}(z)\,dz 
 \Bigg/ \frac{d\Gamma(\BXsll)}{d\hat{s}}.
\eeq
The numerator at the parton level of the forward-backward asymmetries
introduced in (\ref{AFB:unnorm}) and (\ref{AFB:norm}) is 
\begin{align}
 & \int_{-1}^1\frac{d^2\Gamma(\bsll)}{d\hat{s}\,dz}\,
   {\rm{sgn}}(z)\,dz = \left(\frac{\alE}{4\,\pi}\right)^2
   \frac{G_F^2\,(m_b^{\rm{pole}})^5\left|V_{ts}^*V_{tb}^{}\right|^2}
   {48\,\pi^3}(1-\hat{s})^2 \nnl[1mm]
%%%
 & \hspace{3cm} \times \left[ -3\,\hat{s}\,{\rm{Re}}(\widetilde C_9^{\rm{eff}}
   \widetilde C_{10}^{\rm{eff}*})\,
   \left(1+\frac{2\alpha_s}{\pi}\,f_{910}(\hat{s})\right) \right. \nnl[1mm]
%%%
 & \hspace{3.7cm} \left.
   - 6\,{\rm{Re}}(\widetilde C_7^{\rm{eff}} \widetilde C_{10}^{\rm{eff}*})\,
   \left(1+\frac{2\alpha_s}{\pi}\,f_{710}(\hat{s})\right) +
   {\cal A}_{\rm FB}^{\rm Brems}(\hat{s}) \right].
\end{align}

There $z \equiv \cos\theta$ and $\theta$ is the angle between the
positively charged lepton and the $b$ quark in the dilepton center of mass
frame. The functions $f_{ij}(\hat{s})$ summarize virtual and real QCD
corrections \cite{GHIY:npb648,ABGH:prd66}. The real QCD corrections ${\cal A}_{\rm
FB}^{\rm Brems}(\hat{s})$ are infrared-finite \cite{AAHP:mpla19} and their
contribution does not exceed $1\%$ in the SM. In the following they will be
neglected. As in the case of the dilepton invariant mass spectrum the
non-perturbative contributions $(\Lqcd/m_b)^n$ have to be added to pass from the
partonic quantity $d^2\Gamma(\bsll)/d\hat{s}\,dz$ to the hadronic quantity
$d^2\Gamma(\BXsll)/d\hat{s}\,dz$. They can be found in 
\cite{AHHM:prd55,BI:npb525,BB:plb469} whereas the $(\Lqcd/m_c)^2$
corrections are given in \cite{BIR:npb511}.

The position of the zero of these asymmetries, $\hat{s}_0$, is of special
interest because it is sensitive to new physics. It has the value of $\hat{s}_0
= 0.162 \pm 0.010$ at NNLO in the SM. However, as a quantity comparable with
experiments one should consider $q_0^2 = m_b^2 \hat{s}_0$. Therefore an
additional uncertainty due to the $b$-quark mass arises. In
\cite{BGGH:jhep0404,GHIY:npb685} the value of $q_0^2$ has been calculated at the
NNLO in the SM yielding $q_0^2 = (3.76 \pm 0.33)\, \GeV^2$ and $q_0^2 = (3.90
\pm 0.25)\, \GeV^2$, respectively, depending on the choice of $m_b$.

%%%%%%%%%%  Numerical analysis  %%%%%%%%%%%%%%%%%%%%%%%%%%%%%%%%%%%%%%%%%%%%%%%%
\renewcommand{\theequation}{\arabic{section}.\arabic{equation}}
\setcounter{equation}{0}

\section{\large Phenomenological Implications 
         \label{sect:num:ana} }

In what follows we will investigate the phenomenological implications of the
MSSM corrections for the branching ratio, the dilepton invariant mass
distribution and the forward-backward asymmetry.

\subsection{MSSM Parameters and Constraints
            \label{sect:51}}

At the present, neither squark masses nor elements of squark mixing matrices
have been measured, thus it is more appropriate to scan over the fundamental
parameters of the MSSM Lagrangian in order to investigate the new physics
effects. The special scenario of the MSSM under consideration has already been
described in Section \ref{sect:MSSM}.

These fundamental parameters determine the masses and mixing matrices of the
``effective MSSM'' sparticle spectrum at the scale $\mu_\gIno$. We
would like to remind the reader that the MSSM parameters in
\cite{BBKU:npb630,BEKU:prd66} refer to the so-called super-CKM basis
\cite{MPR:superCKM} of the scalar superpartners of the SM fermion sector. 
The fundamental parameters of the MSSM relevant in our numerical analysis
are 
\bit
\item the charged Higgs mass $M_H$ and $\tan\beta$ in the Higgs sector,
\item $\mu$ and $M_2$ that parametrize the chargino sector,
\item the gluino mass $M_\gIno \sim \mu_\gIno$,
\item the soft supersymmetry breaking scalar masses $\MM{}{D}{L},\; \MM{}{U}{R}$ of
      left-handed down- and right-handed up-squarks,
\item the soft supersymmetry breaking trilinear couplings $A_U$ of up-squarks,
\eit
with $\MM{}{D}{L}, \MM{}{U}{R}$ and $A_U$ assumed to be real and diagonal matrices.
Due to the
$SU(2)$ gauge invariance $\MM{}{U}{L}$ is related to $\MM{}{D}{L}$, namely
$\MM{2}{U}{L} = \Vckm\MM{2}{D}{L} \Vckm^\dagger$. Thus the up-squark squared
mass matrix cannot be decomposed into three $2\times2$ block-matrices for an 
arbitrary diagonal $\MM{2}{D}{L}$.

The decoupling of the gluino requires that the masses of all other sparticles
should be lighter compared to the gluino mass and consequently effects of order
$M_{\rm sparticle}/M_\gIno$ can be neglected. This provides an upper bound on
the sparticle spectrum which is chosen to be $\sim 600~ \GeV$. Further lower
bounds have to be fulfilled on sparticle masses by direct searches from
\cite{PDG} \bit
\item $M_\cIno \geq 94\; \GeV$ for the chargino masses,
\item $m_{\uSq} \geq 100\; \GeV$ for the 2 lightest up-squarks whereas the
      remaining squarks are required to be heavier than $250\; \GeV$.  
\eit 
Due to the matching of box-diagrams contributing to $\bsll$ the Wilson
coefficients also depend on the masses of sneutrinos. As such contributions are
rather small we fix their masses to be degenerate, with masses $m_{\tilde{\nu}}$
between $100$ and $300\;\GeV$. Also the down-squarks are approximated by a
common mass $\sim (300 - 500)\;\GeV$, as they only appear in the function
$[T_1]^1_{\tilde{q}}$, which effect is negligibly small.

We have chosen a scenario within the MSSM with values of $\tan\beta < 10$ to
avoid the appearance of additional operators which are not present in the SM
operator basis (\ref{phys:op:basis}).

A very important constraint on new physics models is the total inclusive
branching ratio for $\BXsgamma$.
It has been shown within scenarios of the MSSM \cite{CDGG:npb534,BMU:npb567}
that the NLO QCD corrections of one-loop diagrams with virtual sparticles can become
important and comparable to the present experimental uncertainty of $\Br(\BXsgamma)$.
Also a correlation between the $\BXsgamma$ and the $\BXsll$ decays is
obvious because both involve the Wilson coefficient $\tilde{C}_7$.

The issue of theoretical uncertainties in $\Br(\BXsgamma)$ is not settled yet.
Two main points arise
here. First the choice of the renormalization scheme of the charm quark mass
$m_c$ in the 2-loop matrix elements of the four-quark operators is still a large
theoretical uncertainty of $11\%$ \cite{GM:npb611}. It can only be solved by the
calculation of NNLO corrections to $\Br(\BXsgamma)$ as anticipated in
\cite{BGS:prd67::MS:npb683}. The second point is concerned with the model-dependences
entering the results of $\Br(\BXsgamma)$ measurements when extrapolating to the
lower end of the photon energy spectrum in the experimental analysis.
In \cite{J:slacpub9610} a total inclusive branching ratio
$\Br(\BXsgamma) = (3.34 \pm 0.38) \times 10^{-4}$ with a photon energy cut
$E_0>m_b/20$ was quoted. A very recent analysis of the Belle Collaboration
\cite{belle:incl:bsV} uses the full inclusive spectrum between
$1.8<E_\gamma<2.8\, \GeV$, without invoking theoretical models of the
photon-spectrum. The necessity to introduce the photon energy cut in theoretical
calculations in order to avoid model-dependent experimental results was also
raised  very recently in \cite{N:hepph0408179}.
The method proposed there results in larger uncertainties of the theoretical
prediction of the order of $25\%$. 
In our numerical analysis the most recent SM calculations \cite{GM:npb611,BCMU:npb631}
will be used, however with $E_0>m_b/20$, and the rather conservative interval 
\beq 
  2.0 \times 10^{-4} \leq \Br(\BXsgamma) \leq 5.0 \times 10^{-4} 
\eeq 
to show the correlations with the $\BXsll$ observables.

The values of the SM parameters are taken to be as in \cite{BGGH:jhep0404}
throughout the numerical analysis.

\begin{figure}[t]
\begin{center}
\scalebox{0.65}{\includegraphics{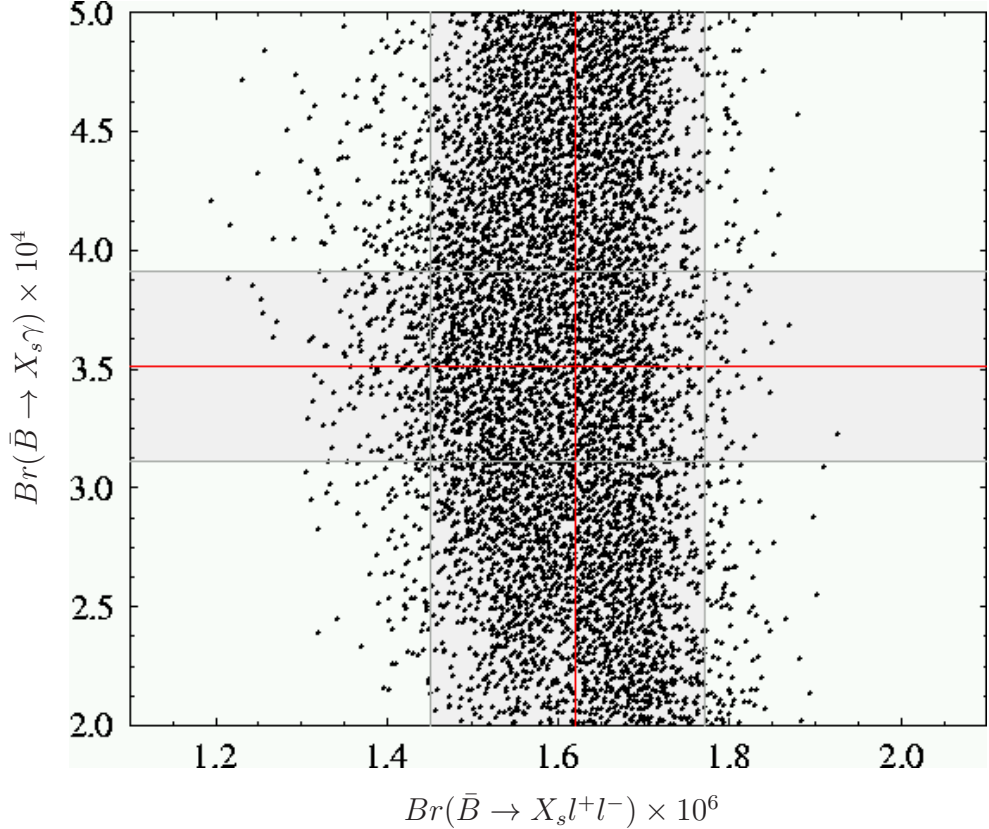}}\\
\hspace*{9mm}\raisebox{-2mm}{$\Br(\BXsll)\times 10^6$}\\
\hspace*{-132mm}
\begin{rotate}{90}\hspace*{45mm}$\Br(\BXsgamma)\times 10^4$\end{rotate}
\caption{\sf $\Br(\BXsgamma)$ versus $\Br(\BXsll)$ for randomly chosen points
  in the parameter space of the MSSM scenario. The three vertical lines indicate
  the SM prediction of $\Br(\BXsll)$ \cite{BGGH:jhep0404} and the 
  three horizontal lines the one for $\Br(\BXsgamma)$ \cite{GM:npb611,BCMU:npb631}. }  
\label{fig:1}
\end{center}
\end{figure}

\subsection{Results}

\begin{figure}[t]
\begin{center}
\begin{tabular}{c@{\hspace{15mm}}c}
\scalebox{0.35}{\includegraphics{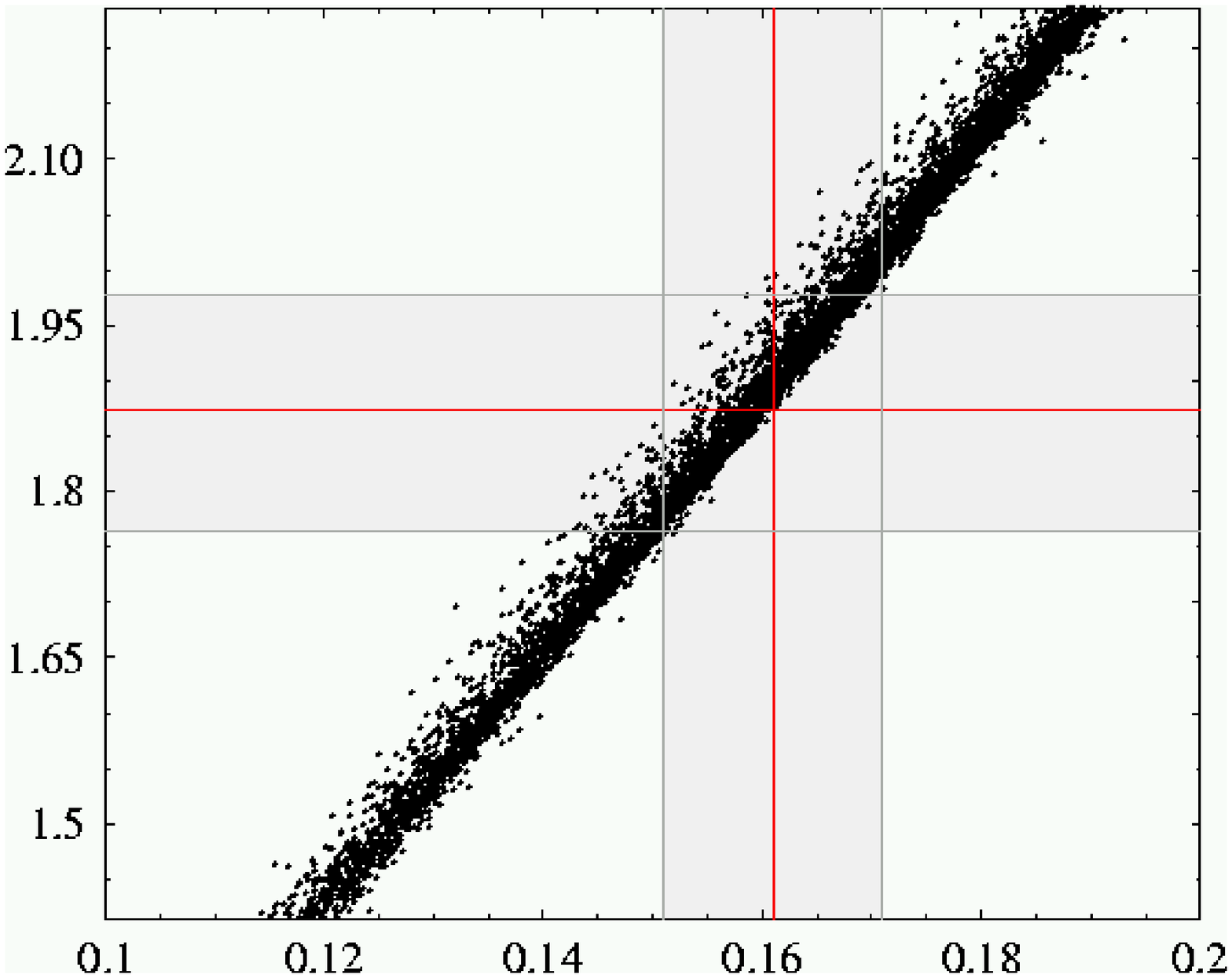}} &
\raisebox{0mm}{\scalebox{0.35}{\includegraphics{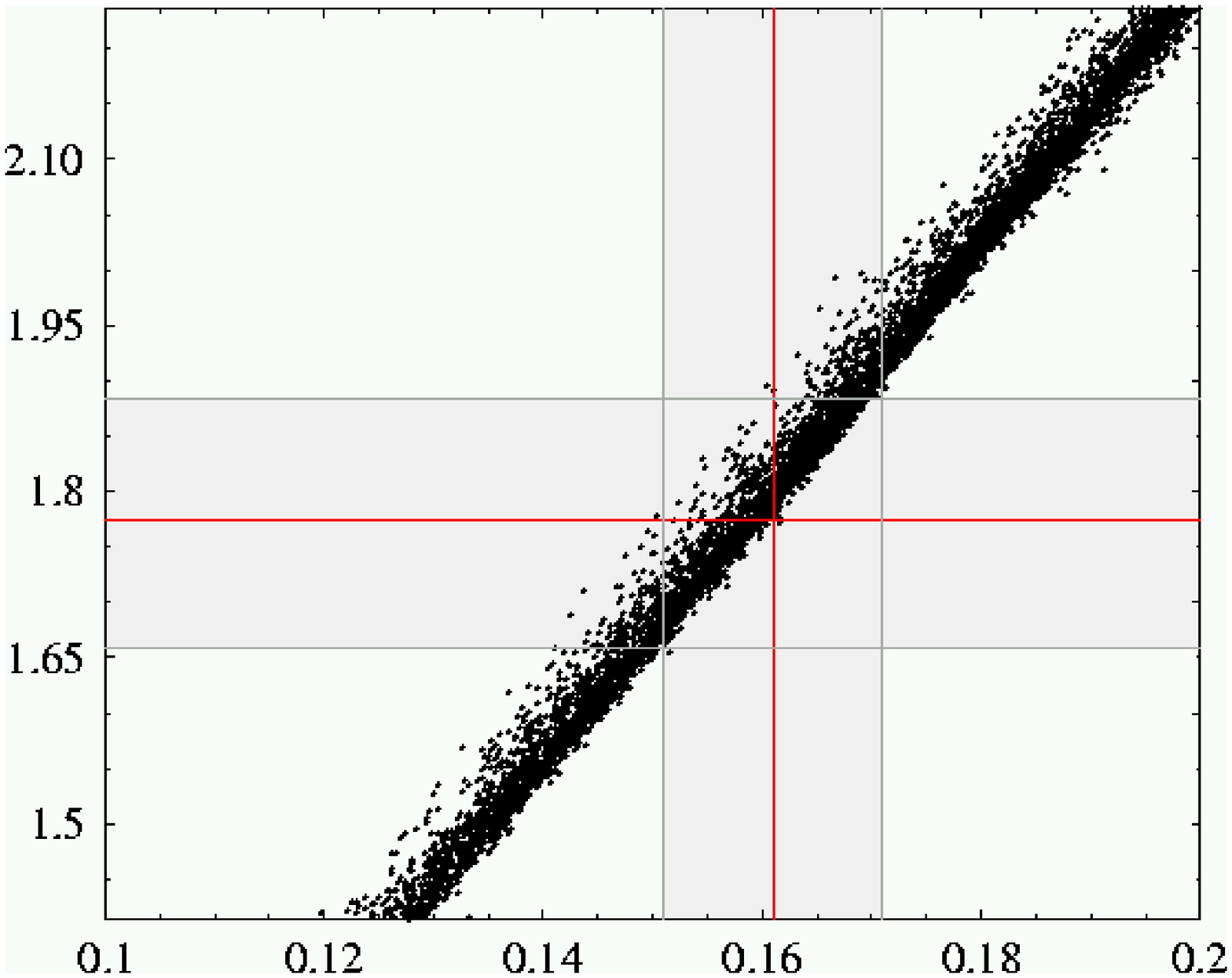}}}\\
\hspace*{6.5mm}\raisebox{-2mm}{\small $\hat{s}_0$} &
 \hspace*{7mm}\raisebox{-2mm}{\small $\hat{s}_0$}\\
\hspace*{-78mm}
\begin{rotate}{90}\hspace*{2cm}\small$\sqrt{Br(\BXsgamma)}\times 10^2$\end{rotate} &
\hspace*{-78mm}
\begin{rotate}{90}\hspace*{2cm}\small$\sqrt{Br(\BXsgamma)}\times 10^2$\end{rotate}
\end{tabular}
\caption{\sf $\sqrt{Br(\BXsgamma)}$ versus $\hat{s}_0$, the position of zero of the
  normalized $\barAFBs$ for randomly chosen points in the parameter space of the
  MSSM scenario. The three vertical lines indicate the SM prediction
  of $\hat{s}_0$ \cite{BGGH:jhep0404,GHIY:npb685} and the three horizontal lines
  the one for $\Br(\BXsgamma)$ \cite{GM:npb611,BCMU:npb631}. 
  In the left plot the $\MSbar$ charm quark mass is used for $\Br(\BXsgamma)$,
  whereas in the right plot $m_c^{\rm pole}$, resulting in a smaller prediction.
}
\label{fig:2}
\end{center}
\end{figure}

\begin{figure}[t]
\begin{center}
\begin{tabular}{c@{\hspace{15mm}}c}
\scalebox{0.35}{\includegraphics{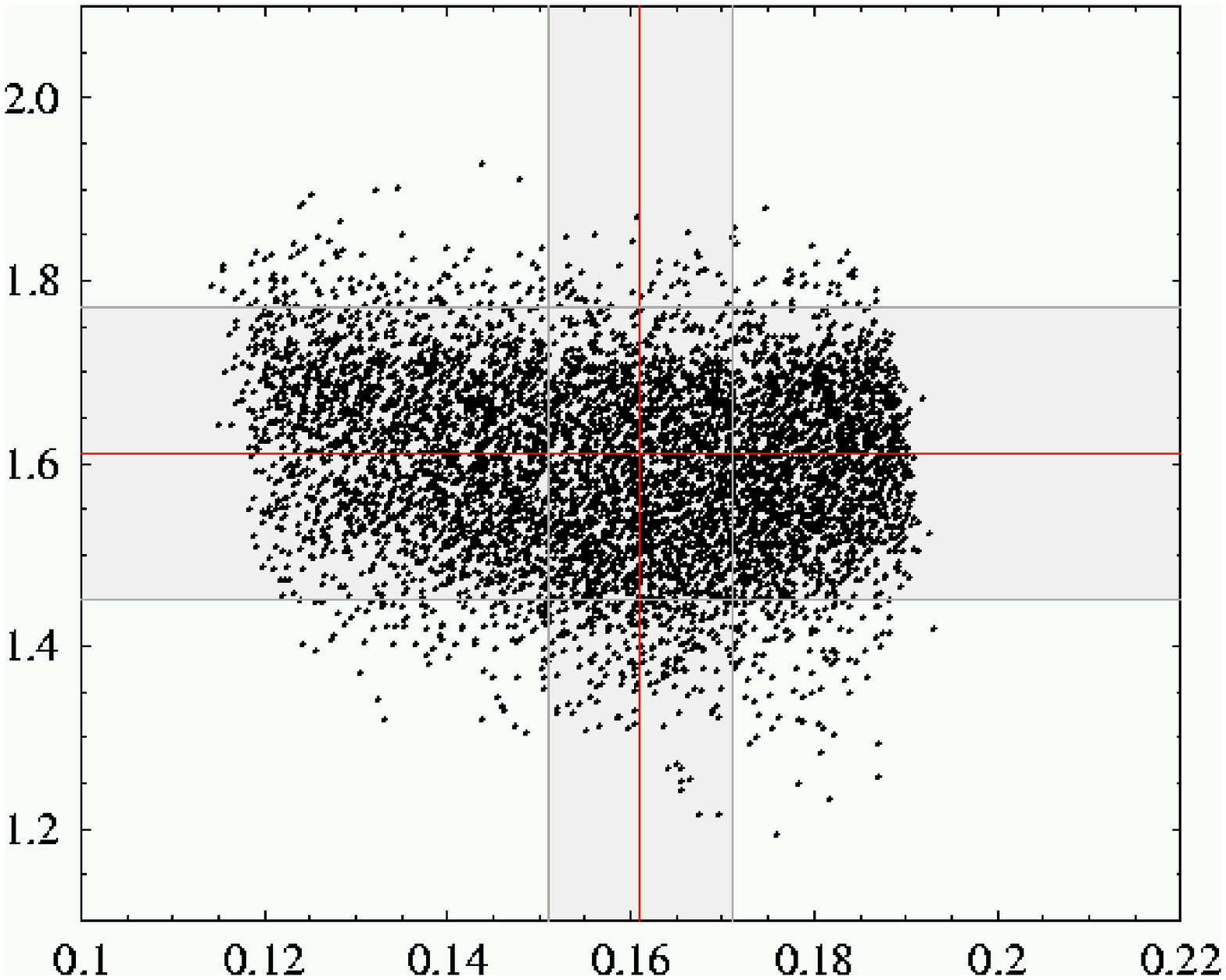}} &
\raisebox{0mm}{\scalebox{0.35}{\includegraphics{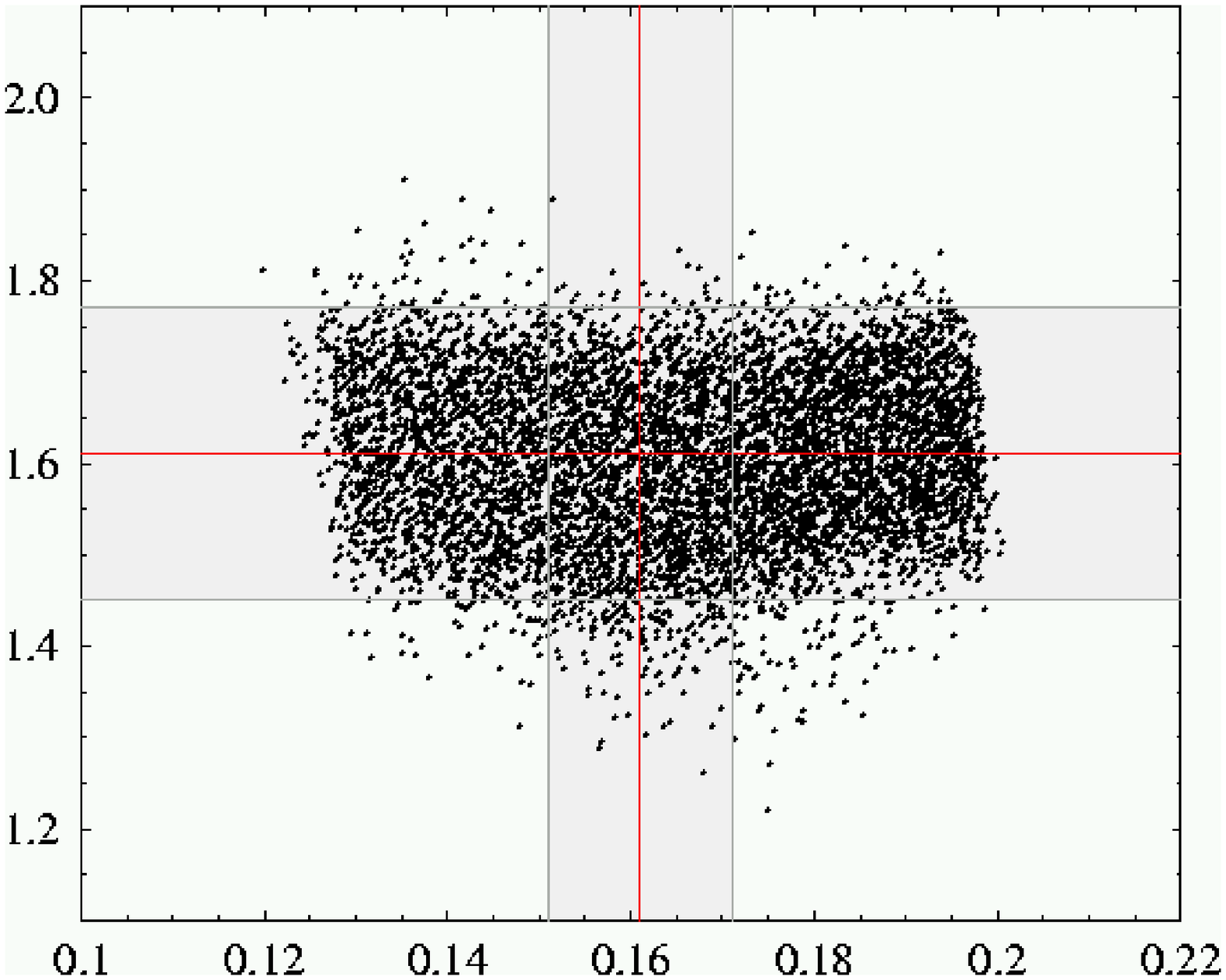}}}\\
\hspace*{6.5mm}\raisebox{-2mm}{\small$\hat{s}_0$} &
 \hspace*{7mm}\raisebox{-2mm}{\small$\hat{s}_0$}\\
\hspace*{-78mm}
\begin{rotate}{90}\hspace*{2cm}\small$Br(\BXsll)\times 10^6$\end{rotate} &
\hspace*{-78mm}
\begin{rotate}{90}\hspace*{2cm}\small$Br(\BXsll)\times 10^6$\end{rotate}
\end{tabular}
\caption{\sf $Br(\BXsll)$ versus $\hat{s}_0$, the position of zero of the
  normalized $\barAFBs$ for randomly chosen points in the parameter space of the
  MSSM scenario. The three vertical lines indicate the SM prediction
  of $\hat{s}_0$ \cite{BGGH:jhep0404,GHIY:npb685} and the three horizontal lines
  the one for $\Br(\BXsll)$ \cite{BGGH:jhep0404}. }
\label{fig:3}
\end{center}
\end{figure}

\begin{figure}[t]
\begin{center}
\begin{tabular}{cc}
  \scalebox{0.58}{\includegraphics{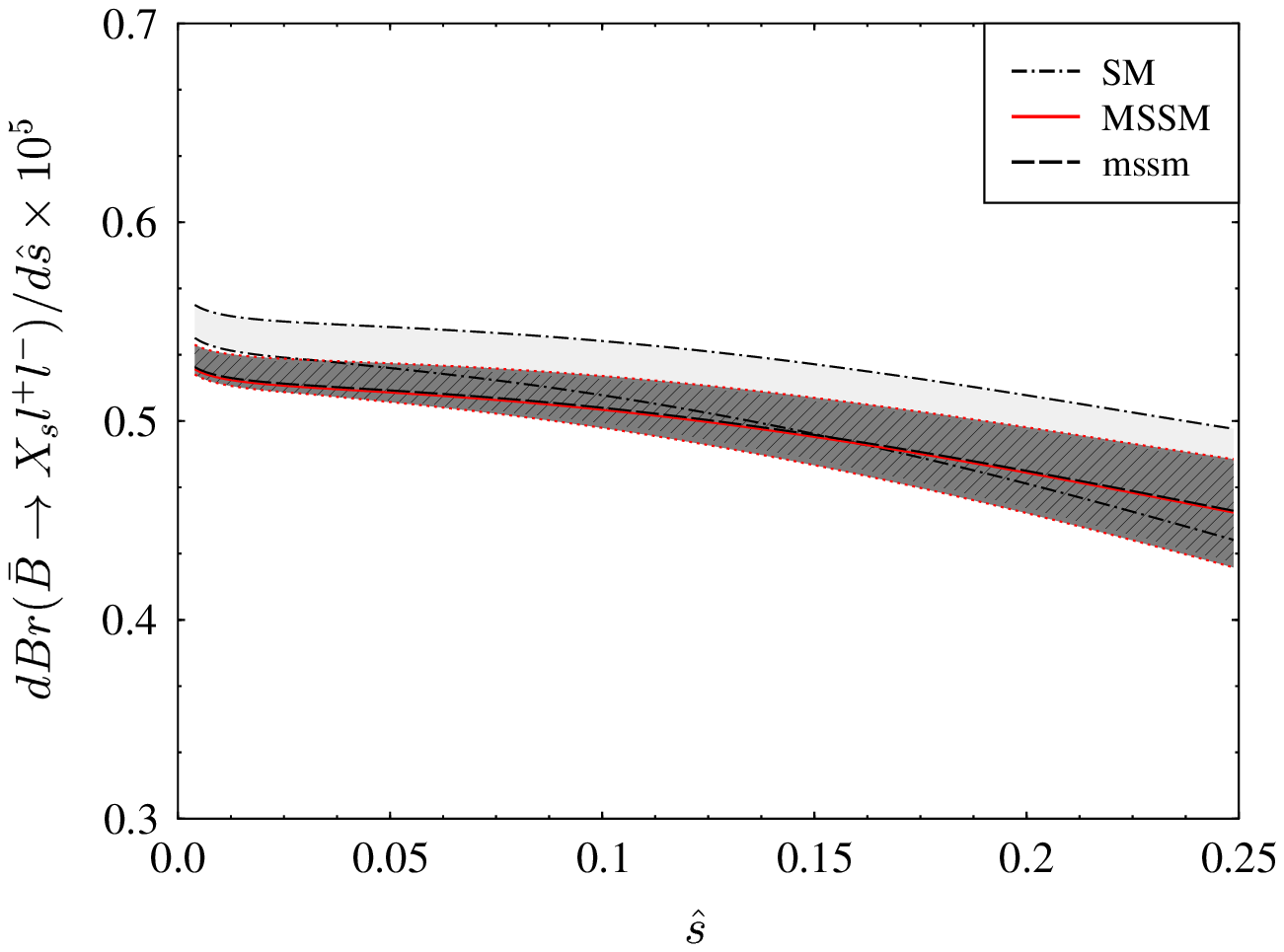}} &
  \scalebox{0.58}{\includegraphics{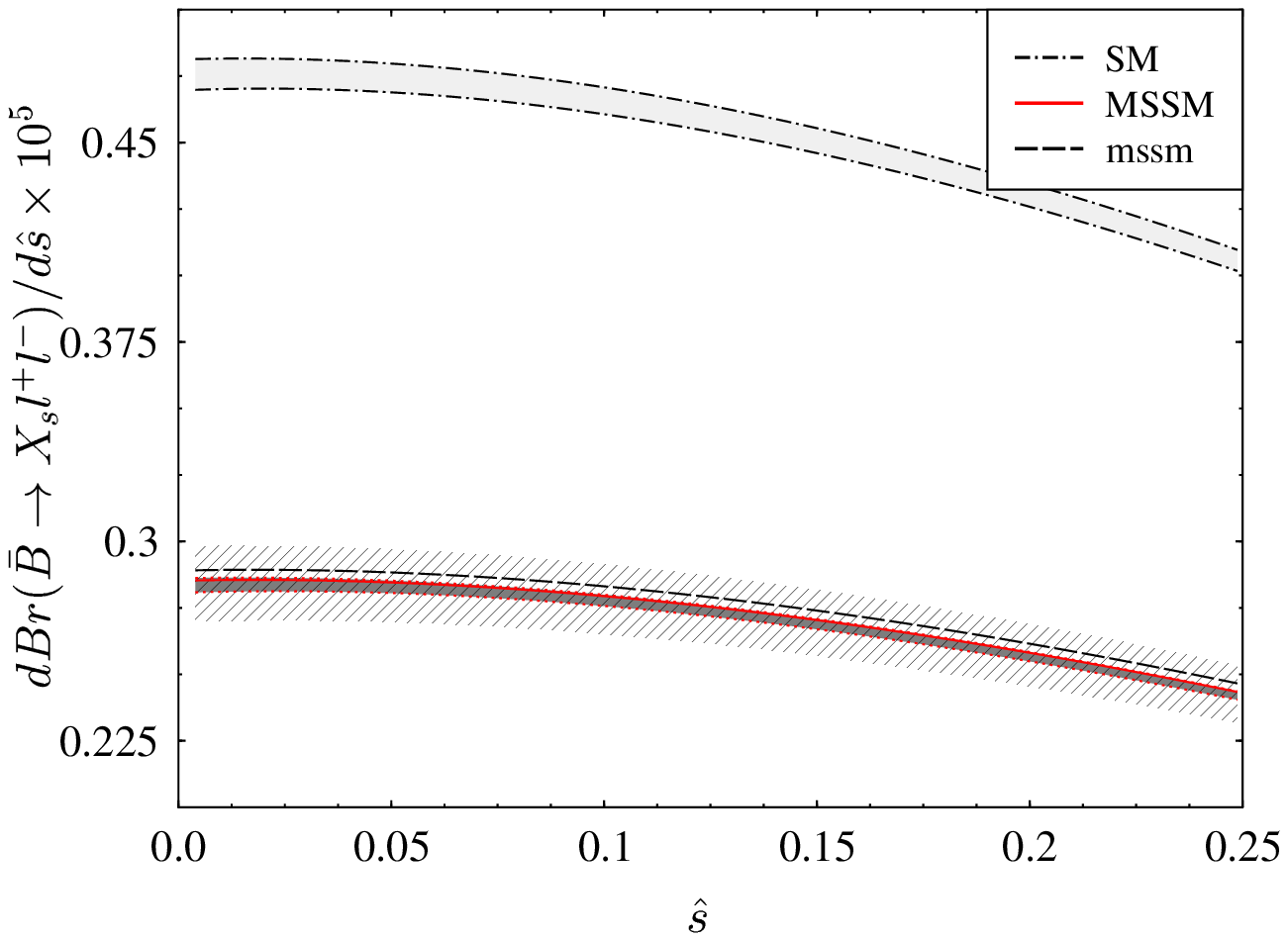}} \\
  $|\widetilde C_9^{\rm{eff}}|^2$ & $|\widetilde C_{10}^{\rm{eff}}|^2$ \\[3mm]
  \scalebox{0.58}{\includegraphics{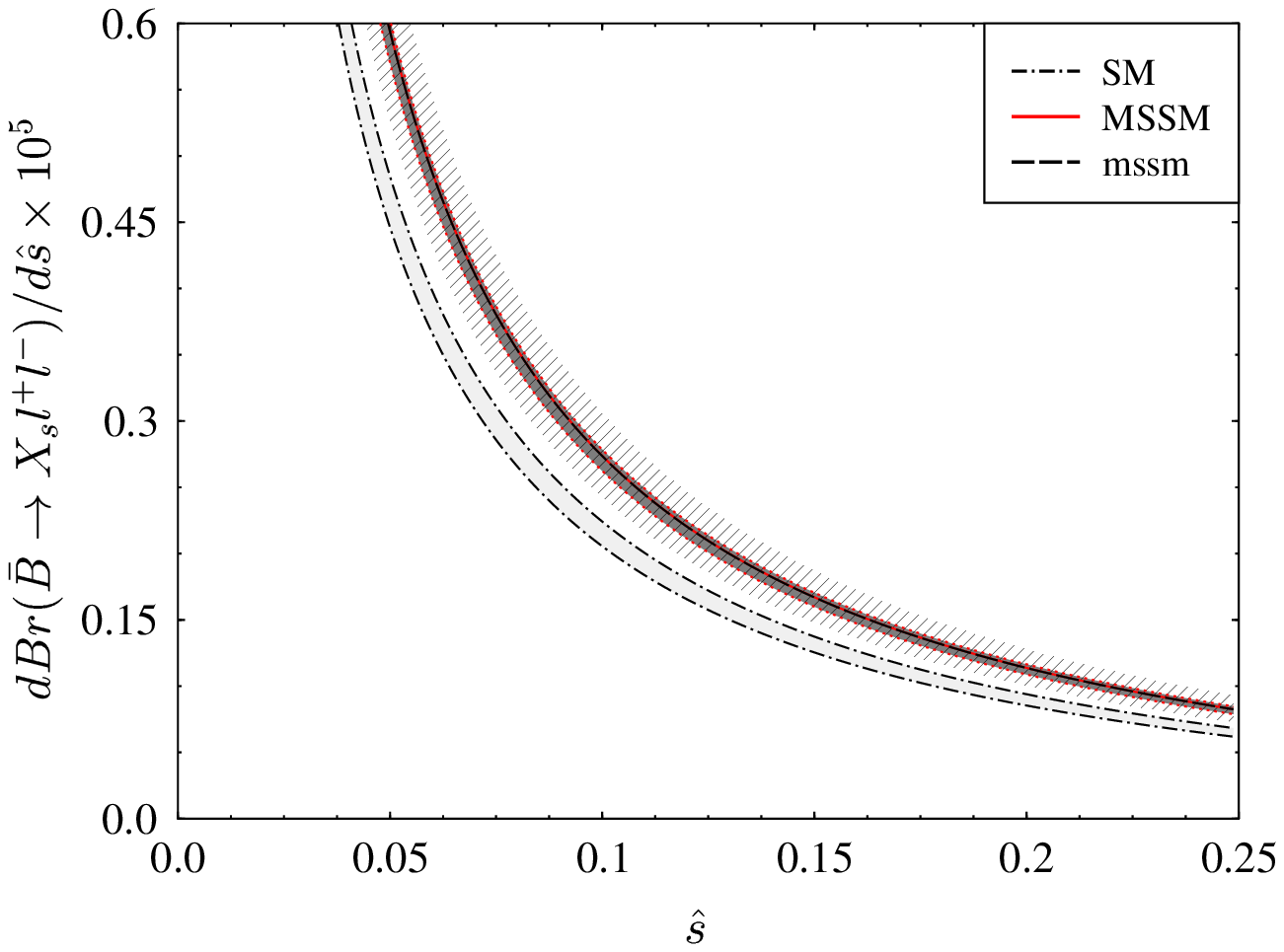}} &
  \scalebox{0.58}{\includegraphics{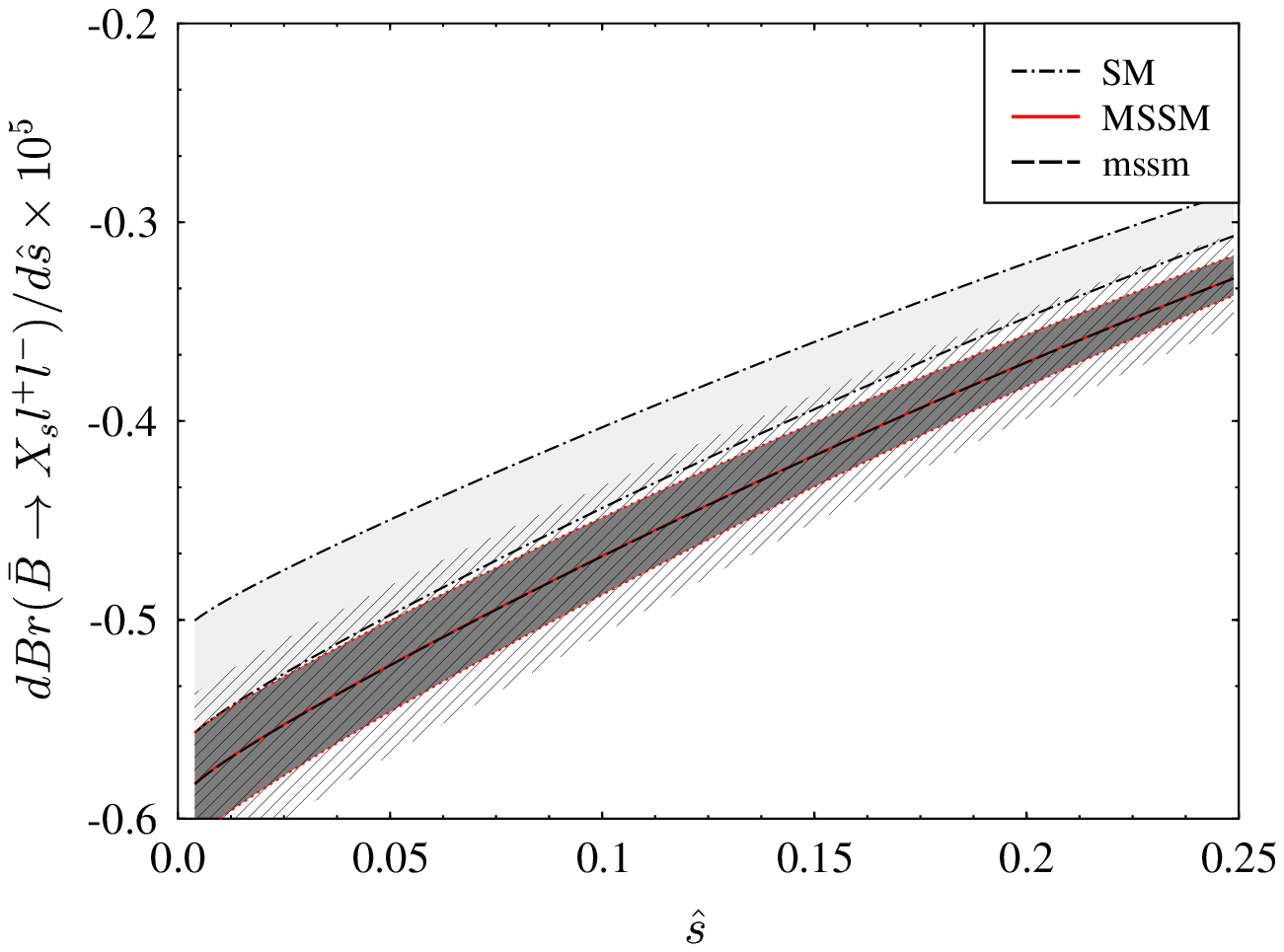}} \\
  $|\widetilde C_7^{\rm{eff}}|^2$ & $\mbox{Re}\left (\widetilde C_7^{{\rm eff}}\widetilde C_9^{{\rm eff}*}\right)$ 
\end{tabular} 
\end{center} 
\caption{\sf Various contributions to the differential branching ratio as 
  functions of $\hat s$. }
\label{fig:4}
\end{figure}

\begin{figure}[t]
\begin{center}
\scalebox{0.58}{\includegraphics{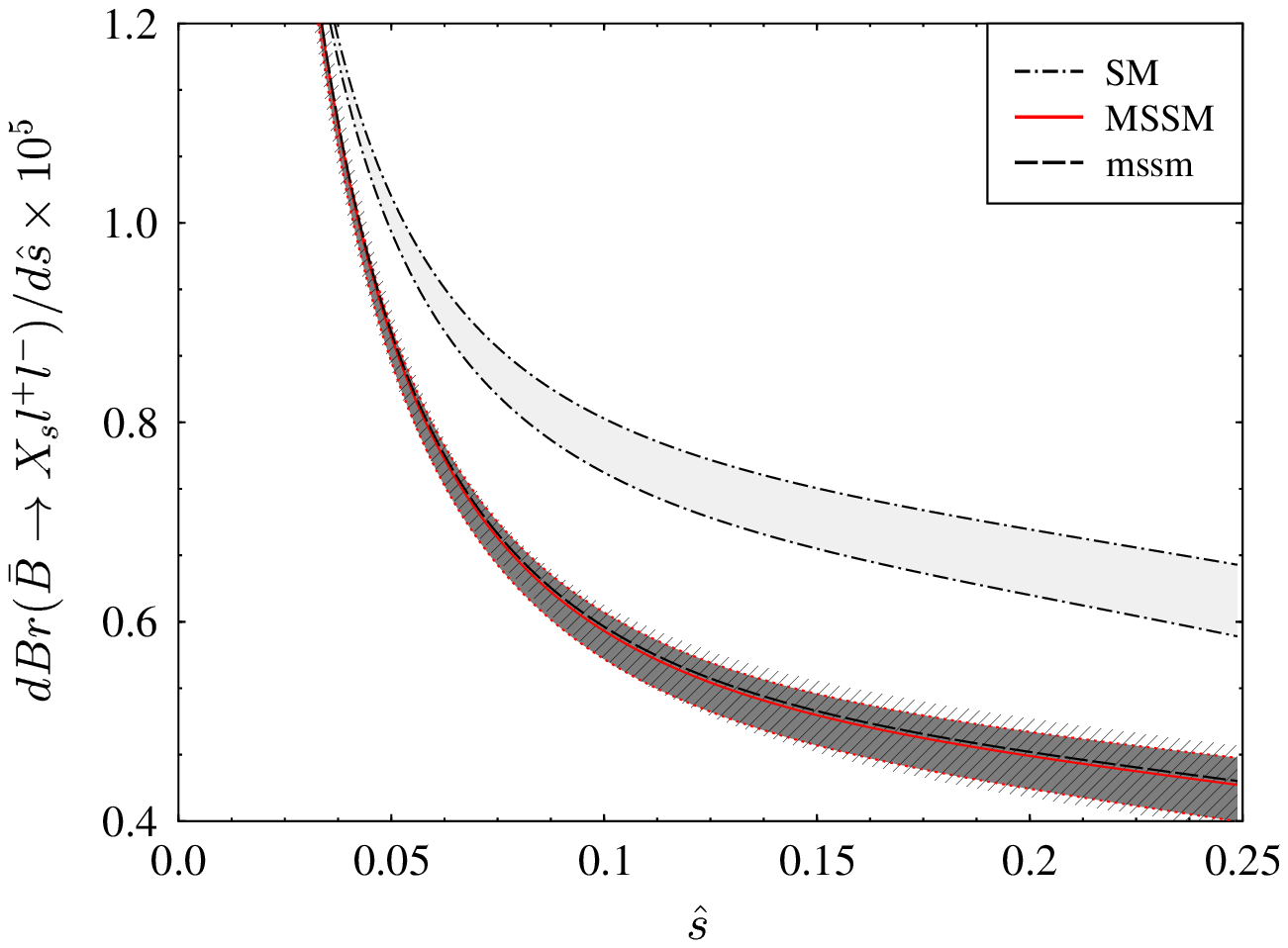}}
\scalebox{0.58}{\includegraphics{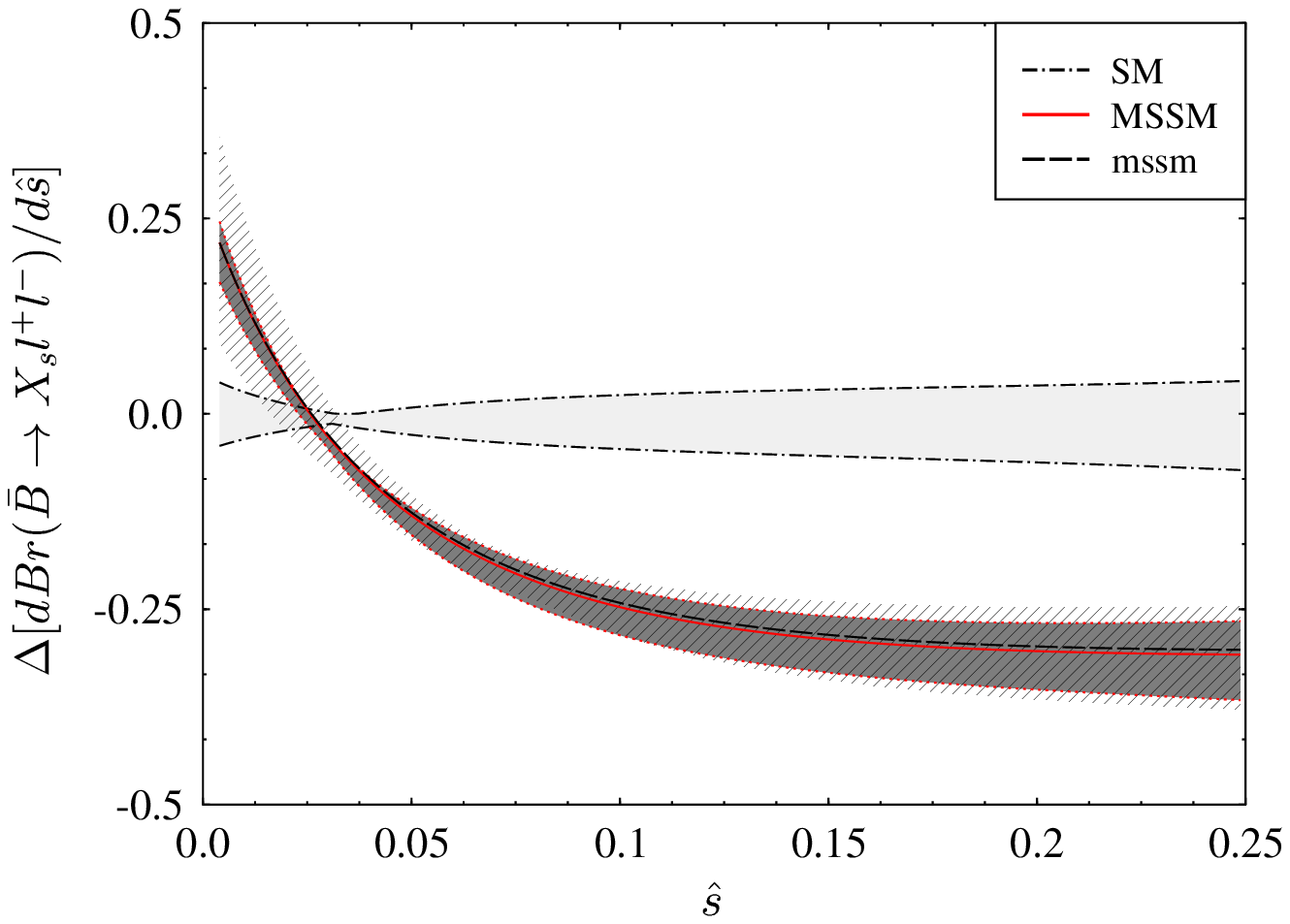}}
\caption{\sf The differential branching ratio for fixed MSSM parameter point
  ``P1'' compared to the SM result and the partial MSSM result as a function of
  $\hat{s}$ (left plot). The relative size compared to the SM is given in the 
  right plot. }
\label{fig:5}
\end{center}
\end{figure}

\begin{figure}[t]
\begin{center}
\scalebox{0.65}{\includegraphics{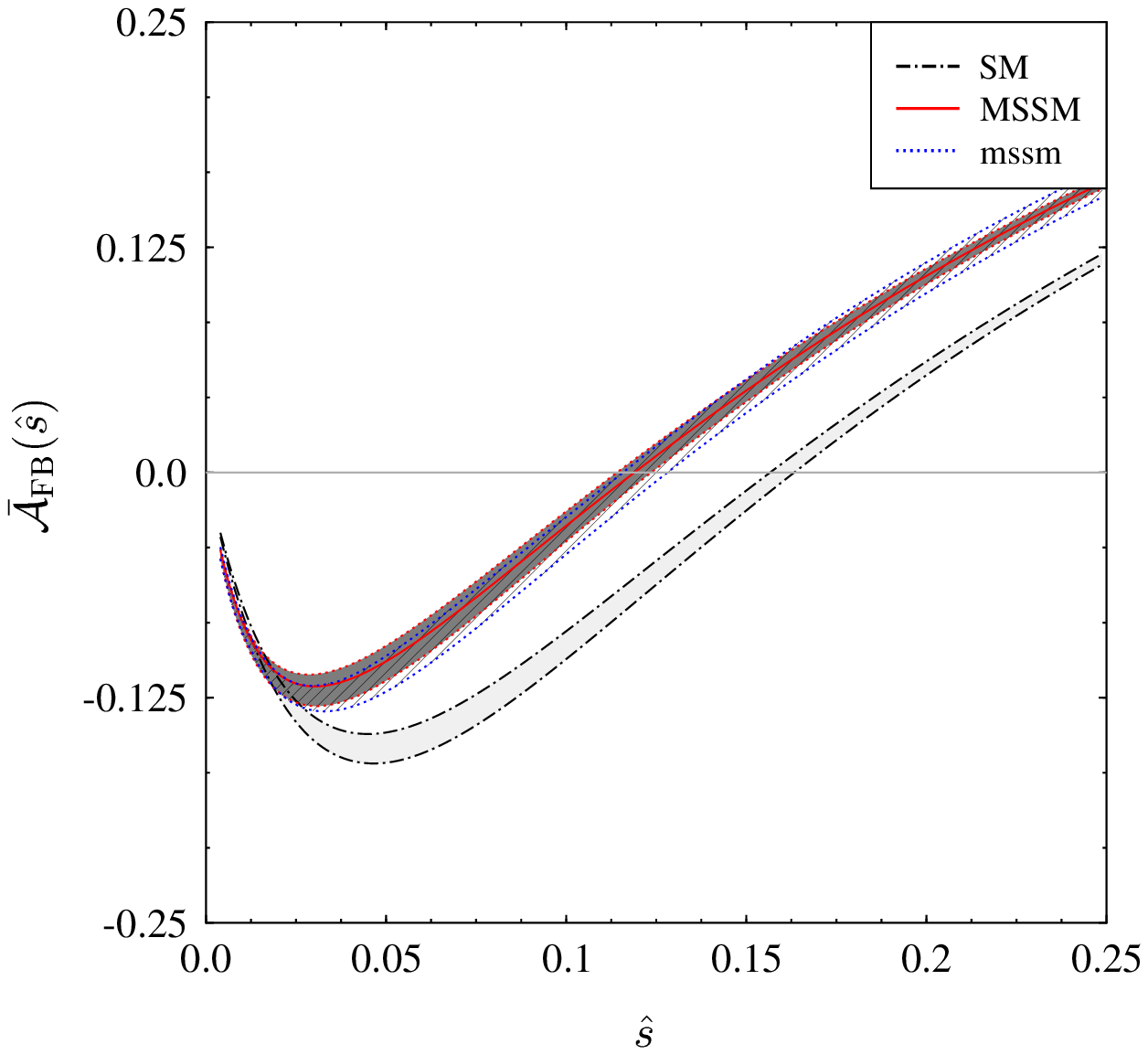}}
\scalebox{0.65}{\includegraphics{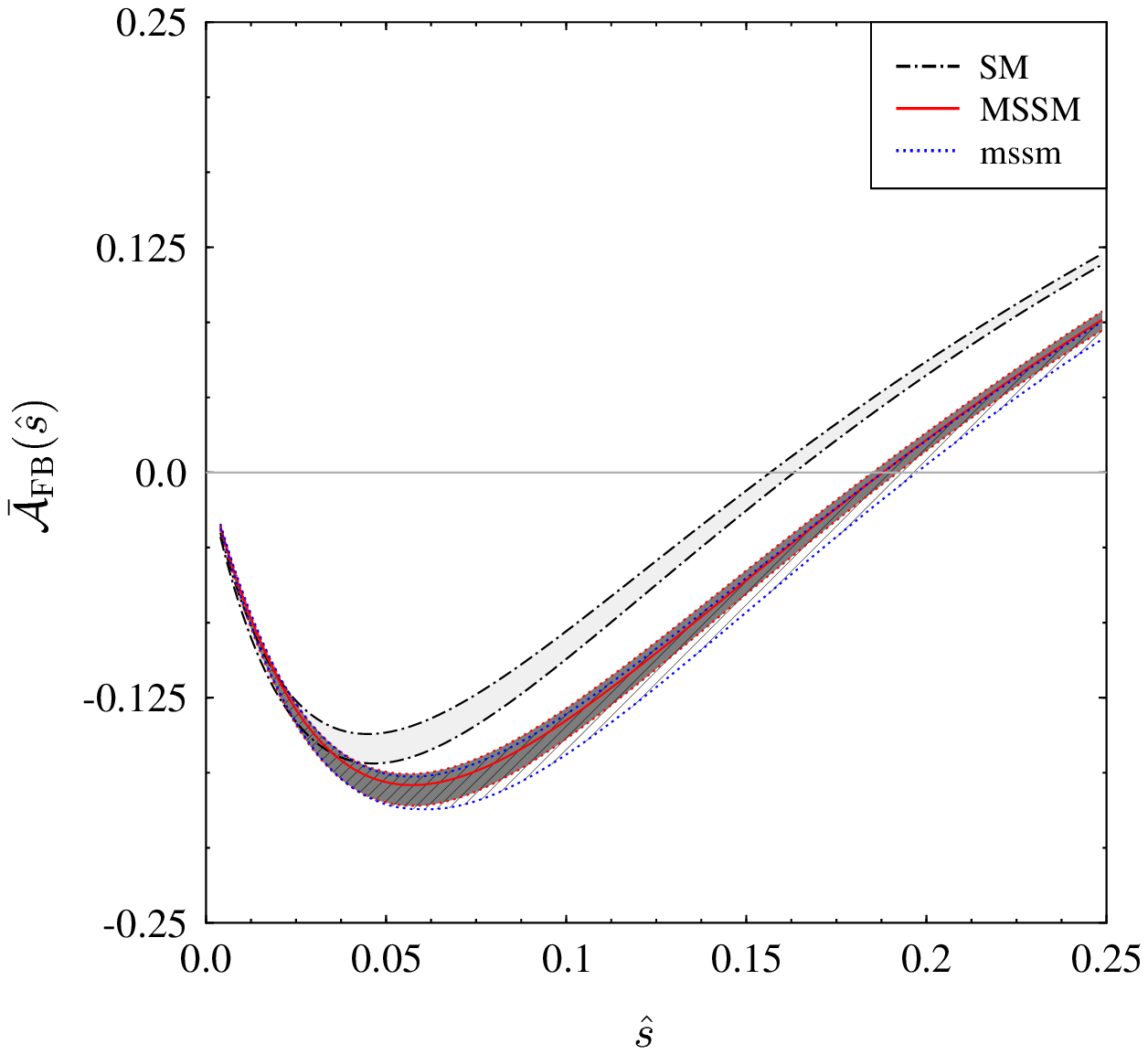}}
\caption{\sf Normalized forward-backward asymmetry $\barAFBs$ versus $\hat{s}$ 
  in the low-$\hat{s}$ region for two (left, right) fixed MSSM-parameter points
  ``P2'' and ``P3'' (see Table~\ref{table:1})  compared with the SM prediction. }
\label{fig:6}
\end{center}
\end{figure}

\begin{figure}[t]
\begin{center}
\begin{tabular}{c@{\hspace{15mm}}c}
\scalebox{0.35}{\includegraphics{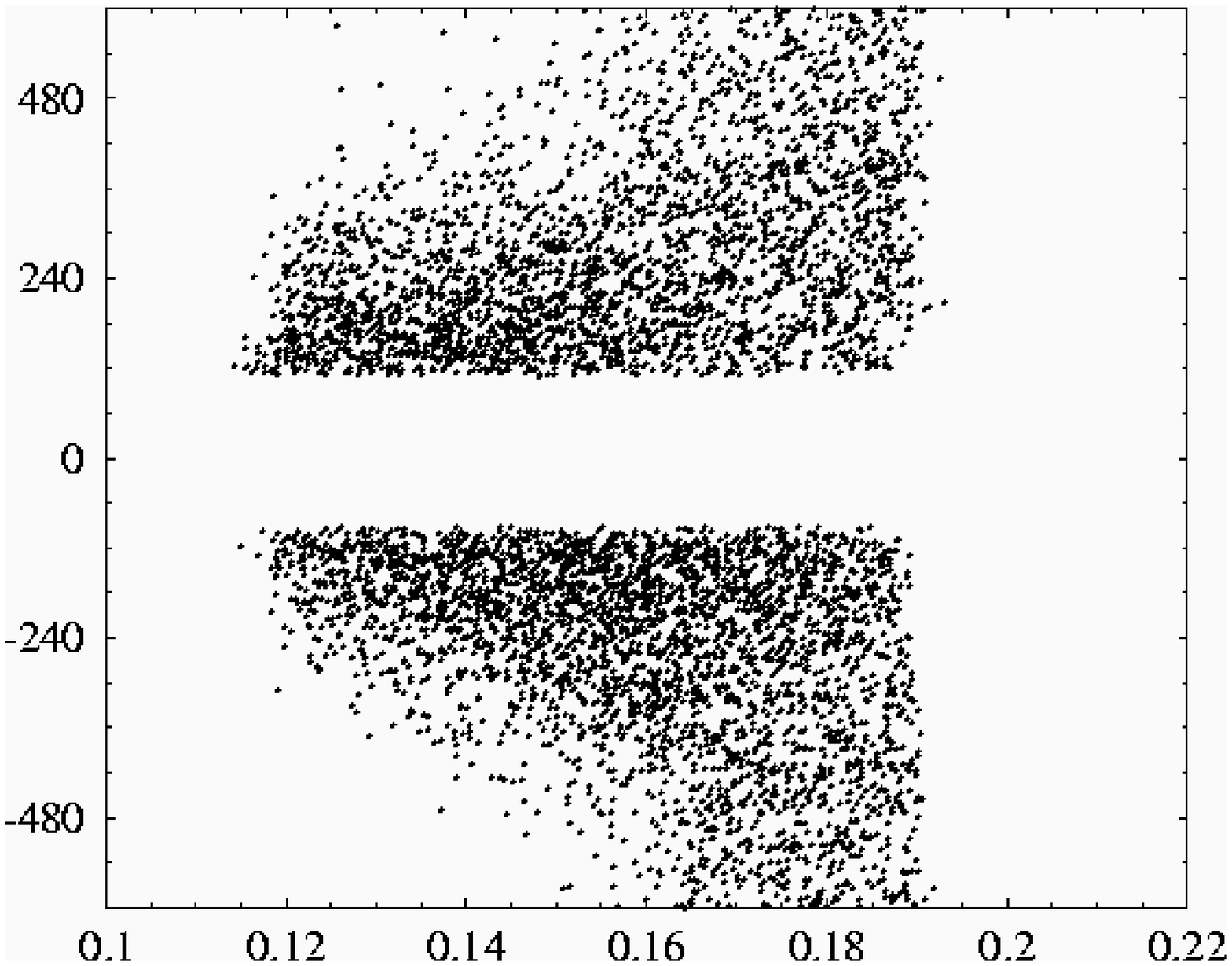}} &
\raisebox{0mm}{\scalebox{0.35}{\includegraphics{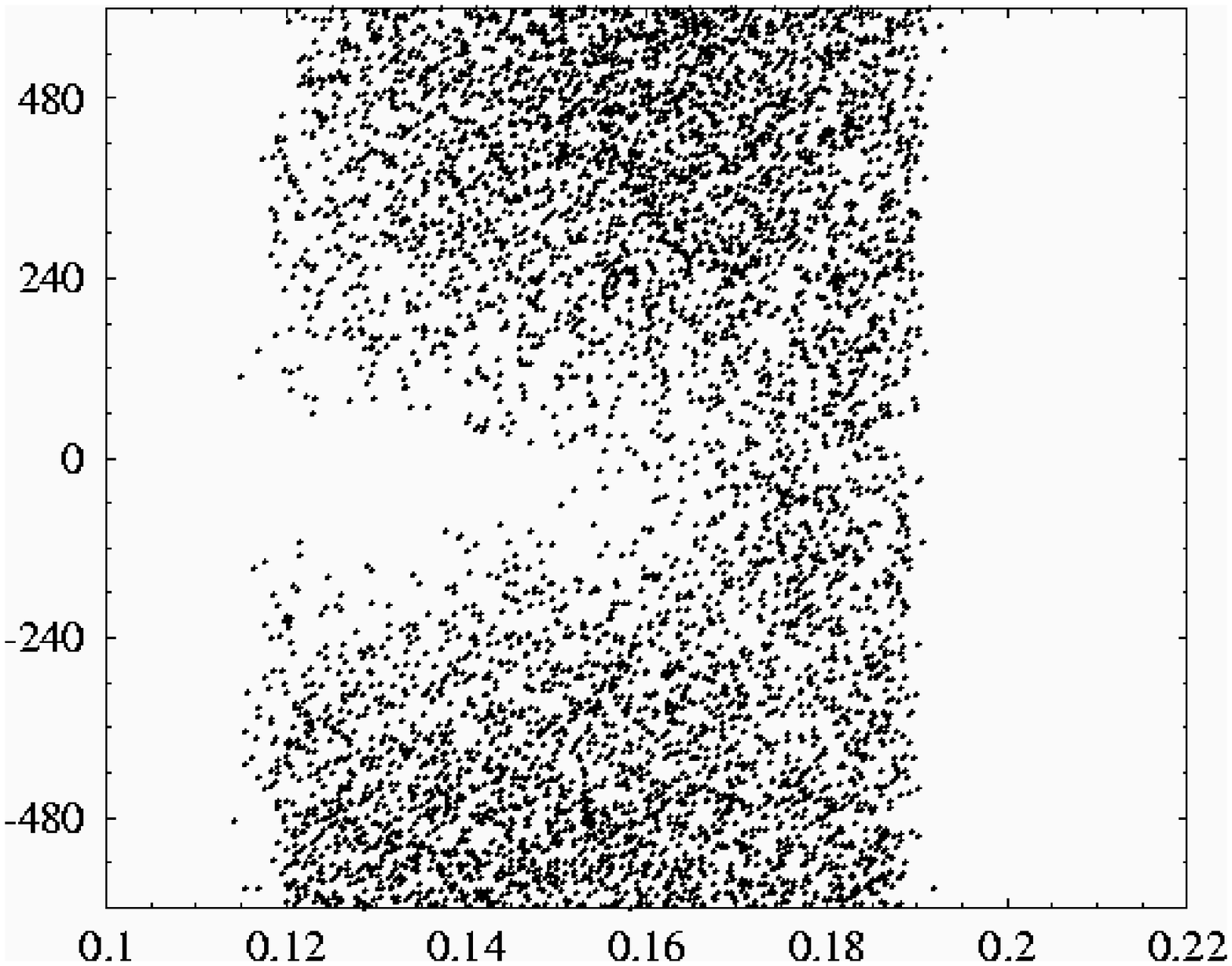}}}\\
\hspace*{7mm}\small$\hat{s}_0$ & \hspace*{7mm}\small$\hat{s}_0$\\
\hspace*{-78mm}
\begin{rotate}{90}\hspace*{38mm}\small$\mu$\end{rotate} &
\hspace*{-78mm}
\begin{rotate}{90}\hspace*{34mm}\small$[A_U]_{33}$\end{rotate}
\end{tabular}
\caption{\sf Correlation between $\hat{s}_0$ and parameters $\mu$ and $[A_U]_{33}$
produced in our scan over MSSM input parameters. The distributions for the other
soft-breaking parameters are flat.}
\label{fig:7}
\end{center}
\end{figure}

We find that the branching ratio $\Br(\BXsll)$ receives only small corrections
within the considered MSSM scenario. This is illustrated in Fig. \ref{fig:1}
where for randomly chosen points of the MSSM parameter space, fulfilling the
lower sparticle mass bounds, the resulting $\Br(\BXsgamma)$ versus $\Br(\BXsll)$
is shown. The vertical lines correspond to the SM prediction of $\Br(\BXsll)$
and the corresponding estimate of the theoretical uncertainty
\cite{BGGH:jhep0404}. The horizontal lines indicate the SM prediction and
theoretical uncertainties of $\Br(\BXsgamma)$ \cite{GM:npb611,
BCMU:npb631}. Deviations are possible from the SM central value $\Br(\BXsll)
\sim 1.60 \times 10^{-6}$ up to $\pm (15-20) \%$ respecting the experimental
bound from $\Br(\BXsgamma)$.  Therefore the observable $\Br(\BXsll)$ of the
low-$s$ region will not serve as a good candidate allowing to distinguish the SM
and the considered MSSM scenario in view of the present theoretical
uncertainties. The reason is the smallness of the MSSM contributions to
$\widetilde C_{9}^{\rm{eff}}$ and $\widetilde C_{10}^{\rm{eff}}$ which dominate
in the expression for $\Br(\BXsll)$ in the low-$s$ region. Although $\widetilde
C_{7}^{\rm{eff}}$ could receive a larger MSSM contribution its magnitude is
strongly constraint by the measured value of $\Br(\BXsgamma)$.\footnote{
  It should be stressed that this is a quite loose
  terminology since for the LO expression of $\Br(\BXsgamma)$ the initial Wilson
  coefficients of the two operators $\Op_7$ and $\Op_8$ enter. At the NLO this
  becomes even more involved. For a model-independent analysis of this subject
  in the presence of new (scalar) operators see \cite{HK:prd69}.}
Furthermore, the contribution to $|\widetilde C_{7}^{\rm{eff}}|^2$ to the
differential branching ratio falls like $1/\hat{s}$ and therefore only dominates
for values of $\hat{s}\lesssim 0.05$ which coincides with the lower end of our
integration range.  The interplay between various contribution to the
differential branching ratio within the SM is depicted in Fig.~\ref{fig:4}.
There also a specific point in the space of supersymmetric parameters with
significant corrections to $\widetilde C_{10}^{\rm{eff}}$ is shown.

The position of the zero of the forward-backward asymmetry $\hat{s}_0$
represents a more sensitive observable than $\Br(\BXsll)$ in the considered MSSM
scenario. In Fig. \ref{fig:2} we plot $\sqrt{\Br(\BXsgamma)}$ versus $\hat s_0$
of the normalized $\barAFBs$ for randomly chosen points of the MSSM parameter
space. There the vertical lines correspond to the SM prediction of $\hat{s}_0$
and its uncertainties \cite{BGGH:jhep0404,GHIY:npb685} and the horizontal lines
as in Fig. \ref{fig:1} to the SM prediction of $\Br(\BXsgamma)$.

We note that the points in both plots in Fig.~\ref{fig:2} are clustered along a
straight line, exhibiting very clearly the correlation between the value of
$Br(B\to X_s\gamma)$ and $\hat s_0$ within models with minimal flavour violation
(MFV) as pointed out in \cite{BPSW:npb678}.

The straight lines in Fig.~\ref{fig:2} are to a very good approximation model
independent within the class of models with MFV. Only different points on them
correspond to different models and/or different sets of parameters in a given
model. On the other hand the position of these lines depends on the parameters
of the low energy theory, in particular on the charm quark mass that enters
sensitively the evaluation of $Br(B\to X_s\gamma)$ \cite{GM:npb611} but is
practically irrelevant for $\hat s_0$. In the left plot in Fig.~\ref{fig:2} we
used the $\MSbar$ mass $\overline{m}_c(m_c)$ and in the right plot the $m_c^{\rm
pole}$ mass, that results in a different straight line.  The SM prediction for
$\Br(\BXsgamma)$ is lower in the right plot than in the left plot.  It is clear
that the usefulness of the correlation between the values of $Br(B\to
X_s\gamma)$ and $\hat s_0$ in testing the MSSM will depend on the progress in
NNLO calculations for $B\to X_s\gamma$ that should significantly decrease the
sensitivity due to the choice of $m_c$.

As seen in Fig.~\ref{fig:2}, in addition to dense points in the ballpark of SM
expectations, there are values of $Br(B\to X_s\gamma)$ and $\hat s_0$ within the
MSSM that are larger and smaller than the SM predictions.  This should be
contrasted with the result in a model with one universal extra dimension in
which only smaller values of $Br(B\to X_s\gamma)$ and $\hat s_0$ were possible
\cite{BPSW:npb678}.

In Fig.~\ref{fig:3} we show $Br(\bar B\to X_s l^+l^-)$ versus $\hat s_0$. In the
left plot the $\MSbar$ definition was used for the charm quark mass in the
evaluation of $\Br(\BXsgamma)$ whereas in the right plot the pole-mass
definition. As a consequence the allowed range of the position of the zero of
$\barAFBs$ becomes shifted a bit towards higher values. The comparison of
Fig.~\ref{fig:2} and \ref{fig:3} shows that the position of $\hat s_0$ is much
more sensitive to the Wilson coefficient $C_7$ and consequently to $Br(B\to
X_s\gamma)$ than to $Br(\bar B\to X_s l^+l^-)$ itself.

\begin{table}[H]
\begin{center}
\begin{tabular}{||l|l||}
\hline \hline
 ``P1'' &  $M_H = 440.11\,\GeV$, $\tan\beta = 5.01$, $\mu = -122.87\,\GeV$, $M_2 = 184.56\,\GeV$,\\
        &  $A_U = {\rm diag}(370.29,\; 79.60,\; 535.71)\,\GeV$, \\
        &  $M_{D_L} = {\rm diag}(299.63,\; 127.20,\; 454.43)\,\GeV$, \\
        &  $M_{U_R} = {\rm diag}(219.96,\; 519.91,\; 167.68)\,\GeV$ \\
\hline
 ``P2'' &  $M_H = 248.34\,\GeV$, $\tan\beta = 2.56$, $\mu = 192.83\,\GeV$, $M_2 = 489.68\,\GeV$,\\
        &  $A_U = {\rm diag}(-419.30,\; 525.64,\; -540.81)\,\GeV$, \\
        &  $M_{D_L} = {\rm diag}(339.09,\; 128.18,\; 393.52)\,\GeV$, \\
        &  $M_{U_R} = {\rm diag}(232.08,\; 351.41,\; 234.77)\,\GeV$ \\
\hline
 ``P3'' & $M_H = 451.74\,\GeV$, $\tan\beta = 4.89$, $\mu = -540.06\,\GeV$, $M_2 = 582.50\,\GeV$,\\
        & $A_U = {\rm diag}(-375.95,\; -324.59,\; -497.23)\,\GeV$, \\
        & $M_{D_L} = {\rm diag}(503.97,\; 281.42,\; 264.06)\,\GeV$, \\
        & $M_{U_R} = {\rm diag}(444.06,\; 186.86,\; 417.40)\,\GeV$. \\
\hline \hline
\end{tabular}
\end{center}
\caption{\sf Three selected points. In all points $M_\gIno = 1\,\TeV$, 
  and the down-squarks and sneutrinos are assumed to have masses about 
  $m_\dSq = 300\,\GeV$ and $m_\Sneu = 250\,\GeV$.}
\label{table:1}
\end{table}

In Fig.~\ref{fig:4} we show the four main contributions due to $|\widetilde
C_{7,9,10}^{\rm{eff}}|^2$ and $\mbox{Re}\left(\widetilde C_7^{{\rm
eff}}\widetilde C_9^{{\rm eff}*}\right)$ to the differential branching ratio,
see (\ref{bsll:diff:Gamma}), as functions of $\hat{s}$ for the fixed MSSM
parameter point ``P1'' defined in Table~\ref{table:1}.  Each plot shows the SM
(light grey band) and the MSSM contribution. To demonstrate the reduction of the
renormalization scale dependence $\mu_t$ we show the MSSM result when including
all calculated corrections (dark grey band -- ``MSSM'') and the partial MSSM
result (shaded bend -- ``mssm'') obtained by discarding all contributions with
$n=2$ and $i=\{H,\cIno,4\}$ to the functions $[X]^n_i$ in (\ref{Xfunctions}),
but not to the SM. The bands are obtained by varying the renormalization scale
$\mu_t \in [120, 300]\, \GeV$ and the low-energy scale $\mu_b \in[2.5, 10] \,
\GeV$. Large deviations from the SM appear in the contribution $|\widetilde
C_{10}^{\rm{eff}}|^2$ mainly due to the $Z^0$-penguin function $[C_9^\lbl]$
which is suppressed in $|\widetilde C_9^{\rm{eff}}|^2$ as can be seen in
(\ref{top:wc}). The inclusion of the NNLO matching conditions in the MSSM
reduces the renormalization scale dependence to comparable size as obtained in
the SM calculation.

The sum of this four separate contributions (and the bremsstrahlung
contributions) adds up to the final differential branching ratio shown in
Fig.~\ref{fig:5} in the left plot. As before the bands are obtained by variation
of the renormalization scales $\mu_t$ and $\mu_b$.  The reduction due to MSSM
contributions is roughly $-30\%$ for values of $\hat{s} > 0.15$ as can be seen
in the right plot of Fig.~\ref{fig:5} where the relative size compared to the SM
result (obtained for $\mu_t = 120\, \GeV$ and $\mu_b = 5 \, \GeV$) is given by
the quantity $\Delta Q \equiv Q/Q_{\rm SM} - 1$.  Thus the shape and magnitude
of the dilepton invariant mass distribution provides in certain regions of
$\hat{s}$ a more sensitive observable then the integrated branching ratio itself
in the search for deviations from the SM prediction, depending on the MSSM
parameter point. It should be noted that the very small scale dependence around
values of $\hat{s} \sim 0.05$ are due to accidental cancellations between the 4
separate contributions in (\ref{bsll:diff:Gamma}).

In Fig.~\ref{fig:6} we show the normalized forward-backward asymmetry $\barAFBs$
for the low-$s$ region. The left plot illustrates the result for the fixed
MSSM-parameter point ``P2'' and the right plot for ``P3'' that are given in
Table~\ref{table:1}. The SM result is shown in both plots for comparison. Again
the bands are obtained by varying the renormalization scales $\mu_t$ and $\mu_b$
as in Figs.~\ref{fig:4} and ~\ref{fig:5}. Due to the strong correlation of the
position of the zero $\hat{s}_0$ and $\Br(\BXsgamma)$ in the considered
MSSM-scenarios further shifts to the left or right (as shown in the two plots)
of $\hat{s}_0$ are unlikely.

In Fig.~\ref{fig:7} the fundamental MSSM parameters $\mu$ and $[A_U]_{33}$ are
shown versus the position of the zero of $\barAFBs$, $\hat{s}_0$, for the sample
of random MSSM points given in Fig.~\ref{fig:2}. The lower and upper bounds of
$\hat{s}_0$ present in both plots are evidently due to the strong correlation to
$\Br(\BXsgamma)$. The ``hole'' in the $\mu$ distribution for values $|\mu| <
100\,\GeV$ comes of course from the bound on the lightest chargino mass. As can
be seen for small values of $\hat{s}_0$ also smaller values of $\mu$ are
preferred. The allowed values of $[A_U]_{33}$ versus $\hat{s}_0$ generated
during our random scan are shown in the right plot. Almost no bounds are found
here, only towards smaller values of $\hat{s}_0$ very small values of
$[A_U]_{33}$ seem to be excluded. We could not find such correlations for all
other soft-SUSY breaking parameters.

%%%%%%%%%%  summary and conclusions  %%%%%%%%%%%%%%%%%%%%%%%%%%%%%%%%%%%%%%%%%%%
\renewcommand{\theequation}{\arabic{section}.\arabic{equation}}
\setcounter{equation}{0}

\section{\large Summary and Conclusions}

In this paper we have presented for the first time complete NNLO QCD corrections
to $\BXsll$ in the context of the MSSM within a scenario as defined in Section
2.  We have calculated the missing ingredients of a complete NNLO result and
including also contributions present already in the literature, we were able to
calculate with this accuracy the branching ratio for $\BXsll$ in the low-$s$
region, the corresponding dilepton invariant mass distribution and the
forward-backward asymmetry. The presented results can be applied to all MSSM
scenarios with a flavour-diagonal down-squark mass-squared matrix, as long as
the gluino is heavier compared to the remaining sparticle spectrum and
$\tan\beta$ is small.

This calculation was motivated by the fact that in the SM the $\BXsll$
observables suffer from sizable renormalization scale uncertainties which are
reduced considerably at NNLO. Consequently in order to have a chance to see
supersymmetric effects of in this decay, it is essential to reduce
renormalization scale uncertainties in the MSSM as well.

The results for the relevant Wilson coefficients are collected in Section 3 and
the Appendix A, where we stated explicitly which corrections were already
calculated previously and which are new. The numerical analysis of the
quantities of interest is performed in Section 5. Our main findings are as
follows:
\begin{itemize}
\item
The $\mu$ dependence present in all quantities of interest at NLO is visibly
reduced at NNLO depending on the magnitude of the MSSM contribution for the
particular MSSM parameter point and it is typically of the same size as the one
of the corresponding SM result.
\item
Supersymmetric effects in the branching ratio amount only to at most $20\%$ and
consequently in view of theoretical uncertainties in this quantity it will be
very difficult to see them unless experimental and theoretical uncertainties
will be significantly reduced.  In this respect the dilepton invariant mass
distribution can offer in certain regions of $\hat{s}$ the possibility to
distinguish the supersymmetric effects from the SM prediction. Such effects can
reach up to $30\%$ depending on the MSSM parameters and the value of $\hat{s}$.
\item
The best chance to observe supersymmetric effects in this decays is through the
forward-backward asymmetry. We find that the position of the zero $\hat{s}_0$ in
this asymmetry can be significantly shifted both downwards and upwards
relatively to the SM expectation. These shifts are accompanied by the shifts in
$\Br(\BXsgamma)$ as pointed out in \cite{BPSW:npb678} and shown in
Fig.~\ref{fig:2}.  As the predictions for $\hat{s}_0$ is theoretically rather
clean, accurate measurements could be able to detect possible departure of its
value from the SM prediction one day.
\end{itemize}

\newpage

%%%%%%%%%%  acknowledgements  %%%%%%%%%%%%%%%%%%%%%%%%%%%%%%%%%%%%%%%%%%%%%%%%%%
\noindent{\large\bf Acknowledgments}

\noindent 
We thank J{\"o}rg Urban, Michael Spranger, Janusz Rosiek and Mikolaj
Misiak for discussions.
C.B. and T.E. have been supported by the German-Israeli Foundation under the
contract G-698-22.7/2002.
Further C.B. has been supported by the Department of Energy under Grant DE-FG03-97ER40546.
The work presented here was also supported in part by the German 
Bundesministerium f{\"u}r Bildung und Forschung under the contract 05HT4WOA/3 and
DFG Project Bu.\ 706/1-2.

%%%%%%%%%%  apppendix  %%%%%%%%%%%%%%%%%%%%%%%%%%%%%%%%%%%%%%%%%%%%%%%%%%%%%%%%%

\renewcommand{\thesection}{Appendix~\Alph{section}}
\renewcommand{\theequation}{\Alph{section}.\arabic{equation}}

\setcounter{equation}{0}
\setcounter{section}{0}

%%%%%%%%%  app. wilson coefficients  %%%%%%%%%%%%%%%%%%%%%%%%%%%%%%%%%%%%%%%%%%%
\section{\large Wilson Coefficients}\label{app:wil:coeff}

This appendix summarizes the matching results relevant for $\BXsll$ in the SM and
the considered scenario of the MSSM as introduced in Section \ref{sect:MSSM}. It
provides the formulae for the functions $[X]^n_i$ introduced in (\ref{Xfunctions}).

Dimensional regularization with fully anticommuting $\gamma_5$ and the $\MSbar$
scheme was used for all QCD counterterms, both in the full and in the effective
theory for light degrees of freedom. The only exceptions were the top quark and
squark loop contributions to the renormalization of the light-quark and gluon
wave functions on the full theory side. The corresponding terms in the
propagators were subtracted in the MOM scheme at $q^2=0$. In consequence, no top
quark and squark loop contribution remained in the ``light quark -- $W$ boson''
effective vertex after renormalization.

The only relevant off-shell electroweak counterterm in the full theory
proportional to $\bar{s} D\!\!\!\!/\,b$ was taken in the MOM scheme as well, at
$q^2=0$ for the $\bar{s} \partial\!\!\!/b$, and at vanishing external momenta
for terms containing gauge bosons.

As a consequence of this special choice of the renormalization, all masses of
quarks and squarks, as well as the mixing matrix $\Gamma^U$ and the effective
couplings $X_{i}^{U_L}$ and $X_{i}^{U_R}$ appearing in this Appendix are
$\MSbar$ quantities.  The masses of particles which do not interact strongly are
not renormalized and thus might be interpreted as their tree-level masses.

As already stated in Section \ref{sect:wil:coeff} all the functions $[X]^1_4$
are equal to zero choosing an on-shell renormalization prescription for squark
fields and masses \cite{Aoki:etal} and the mixing matrix $\Gamma^U$
\cite{Y:prd64}. For example this can be seen by means of the following
transformation formulae between $\MSbar$ and on-shell scheme,
\begin{align}
  \quad m^2_{\uSq_a}(\mu) & = 
    (m^{\rm pole}_{\uSq_a})^2 \Bigg\{1- \frac{\alS(\mu)}{4\pi} \frac{4}{3} 
    \left[ 7 + 3 \ln \Bigg(\frac{\mu}{m^{\rm pole}_{\uSq_a}}\Bigg)^2 \right]
  \nnl[1mm]
  \label{MS:OS:mass:shift}
    & \hspace{2.5cm} + \frac{\alS(\mu)}{4\pi} \frac{4}{3} \sum_{b=1}^6
      \frac{P^U_{ab} (m^{\rm pole}_{\uSq_b})^2 P^U_{ba}}{(m^{\rm pole}_{\uSq_a})^2}
      \left[1 + \ln \Bigg(\frac{\mu}{m^{\rm pole}_{\uSq_b}} \Bigg)^2 \right]\Bigg\},
  \\[2mm]
  \label{MS:OS:GamU:shift}
  \Gamma^{U}_{ab}(\mu) &= \Gamma^{U}_{ab}
      + \frac{\alS(\mu)}{4\pi} \frac{4}{3} 
        \sum_{e=1}^6 \sum_{\substack{\scriptstyle c=1 \\ \scriptstyle c\neq a}}^6
        \frac{P^U_{ae} (m^{\rm pole}_{\uSq_e})^2 P^U_{ec}}
             {(m^{\rm pole}_{\uSq_a})^2-(m^{\rm pole}_{\uSq_c})^2}
        \left[1 + \ln \Bigg(\frac{\mu}{m^{\rm pole}_{\uSq_e}}\Bigg)^2\right]
       \Gamma^{U}_{cb}.
\end{align}

On the left hand side of these equations the squark masses and the mixing matrix
are running $\MSbar$ parameters, whereas on the right hand side they take their
on-shell values. We note that the couplings $a_g$ and $a_Y$ given in
(\ref{eff:gluino}) are still $\MSbar$ renormalized
working with on-shell squark parameters.

We define the mass ratios 
\begin{align}
  x &= \frac{m_t^2}{M_W^2}, & 
  y &= \frac{m_t^2}{M_H^2}, &
  x_{ij} &= \frac{M_{\cIno_i}^2}{M_{\cIno_j}^2}, &
  y_{ai} &= \frac{m_{\uSq_a}^2}{M_{\cIno_i}^2},  & 
  v_{fi} &= \frac{m_{\tilde\nu_f}^2}{M_{\cIno_i}^2},
\end{align}
and introduce the abbreviations
\begin{align}
  L_t & = \ln\frac{\mu_t^2}{m_t^2}, &
  L_{\uSq_a} & = \ln\frac{\mu_t^2}{m_{\uSq_a}^2}, &
  \kappa & = \frac{1}{g_2^2 V_{tb}^{} V_{ts}^{\ast}}.
\end{align}
In these equations $m_t$ denotes the top quark mass, $m_{\uSq_a}$ and
$m_{\dSq_a}$ up and down squark masses, $M_W$ the $W$ boson mass, $M_H$ the
charged Higgs mass, $M_{\cIno_i}$ the chargino masses and finally
$m_{\tilde\nu_f}$ sneutrino masses.

The integral representations for the functions $\Li{z}$ and $\Cl{x}$ are as
follows
\begin{align}
  \Li{z} &= -\int_0^z dt \frac{\ln (1-t)}{t}, 
\\[3mm]
  \Cl{x} &= {\rm Im}\left[ \Li{e^{ix}} \right] =
  -\int_0^x d\theta \ln |2 \sin(\theta/2)|.
\end{align}

The calculation was performed in the background field formalism in an arbitrary
$R_\xi$-gauge for the gluon gauge parameter and in the t`Hooft-Feynman gauge for
the $W$ boson gauge parameter.

%
%
%%%%%%%%%%%%%%%%%%%%
\subsection{\boldmath{$i=W$} -- ``top quark -- W boson''}

The evaluation of Feynman diagrams contributing to $b\to s+$(light particles)
Greens functions within the SM mediated by ``top quark -- $W$ boson'' loops
yields the functions denoted by the index $i=W$ in (\ref{Xfunctions}). The
explicit form can be found in \cite{BMU:npb574} by using the equalities
\beq
  \begin{array}{ccccc}
  {[A_7]}^0_W = A_0^t(x), & \quad &
  {[B_9^\lbl]}^0_W = {[B_{10}^\lbl]}^0_W = B_0^t(x), & \quad &
  {[C_9^\lbl]}^0_W = C_0^t(x), 
\\[4mm]
  {[D_9]}^0_W = D_0^t(x), &&
  {[E_4]}^0_W = E_0^t(x), && 
  {[F_8]}^0_W = F_0^t(x), 
\\[4mm]
  {[A_7]}^1_W = A_1^t(x), &&
  {[B_9^\lbl]}^1_W = {[B_{10}^\lbl]}^1_W = B_1^t(x, -\frac{1}{2}), &&
  {[C_9^\lbl]}^1_W = C_1^t(x), 
\\[4mm]
  {[D_9]}^1_W = D_1^t(x), &&
  {[E_4]}^1_W = E_1^t(x), &&
  {[F_8]}^1_W = F_1^t(x), 
\\[4mm] 
  {[G_3]}^1_W = G_1^t(x). && &&
  \end{array}
\eeq
The functions $[X]^1_W$ have been first calculated in the following papers:
$[A_7]^1_W$ and $[F_8]^1_W$ in \cite{AY:prd49,GH:prd56,CDGG:npb527,BKP:npb517,BMU:npb567}, 
$[B_9^\lbl]^1_W$ and $[C_9^\lbl]^1_W$ in \cite{BB:npb398,BB:npb400,BB:npb548,MU:pl451} and
$[D_9]^1_W,\, [E_4]^1_W$ and $[G_3]^1_W$ in \cite{BMU:npb574}.

%
%
%%%%%%%%%%%%%%%%%%%%%%%%%%%%%%
\subsection{\boldmath{$i=H$} -- ``top quark -- charged Higgs''}

The evaluation of Feynman diagrams contributing to $b\to s+$(light particles)
Greens functions within the MSSM (but also 2HDM of type II) mediated by
``top quark -- charged Higgs boson'' loops and denoted by
the index $i=H$ in (\ref{Xfunctions}) yields
\mathindent=0cm
\begin{align}
  {[A_7]}^0_H &= \tSt  
    \frac{-3y^2+2y}{3(y-1)^3} \ln y + \frac{5y^2-3y}{6(y-1)^2} + 
    \cot^2\beta \left\{ \frac{-3y^3+2y^2}{6(y-1)^4} \ln y
    + \frac{8y^3+5y^2-7y}{36(y-1)^3} \right\}, 
\\[6mm]
  {[B_9^\lbl]}^0_H &= {[B_{10}^\lbl]}^0_H  = 0,
\\[6mm] 
  {[C_9^\lbl]}^0_H &= \tSt 
    \frac{M_H^2}{8 M_W^2} \cot^2\beta \left\{
     \frac{-y^2}{(y-1)^2} \ln y + \frac{y^2}{y-1} \right\},
\\[6mm] 
  {[D_9]}^0_H &=  \tSt
    \cot^2\beta \left\{ 
    \frac{-3y^4+6y^2-4y}{18(y-1)^4} \ln y + \frac{47y^3-79y^2+38y}{108(y-1)^3} \right\},
\\[6mm]
  {[E_4]}^0_H &= \tSt  
    \cot^2\beta \left\{ 
    \frac{3y^2-2y}{6(y-1)^4} \ln y + \frac{7y^3-29y^2+16y}{36(y-1)^3} \right\},
\\[6mm]
  {[F_8]}^0_H &= \tSt  
    \frac{y}{(y-1)^3} \ln y + \frac{y^2-3y}{2(y-1)^2} + 
    \cot^2\beta \left\{ \frac{y^2}{2(y-1)^4} \ln y + \frac{y^3-5y^2-2y}{12(y-1)^3} \right\},
\\[6mm]
  {[A_7]}^1_H &= \tSt
    \frac{-64 y^3 + 224 y^2 - 96 y}{9(y - 1)^3} \Li{1 - \frac{1}{y}}
  + \frac{-28 y^3 + 256 y^2 - 132 y}{9(y - 1)^4} \ln y
  + \frac{16 y^3 - 104 y^2 + 56 y}{3(y - 1)^3} 
  \nnl[2mm] & \tSt
  + \left[ \frac{24 y^3 + 112 y^2 - 64 y}{9(y - 1)^4} \ln y 
         + \frac{32 y^3 - 188 y^2 + 84 y}{9(y - 1)^3} \right] L_t 
  \nnl[2mm] & \tSt
 + \cot^2\beta \left\{
    \frac{-32 y^4 + 148 y^3 - 72 y^2}{9(y - 1)^4} \Li{1 - \frac{1}{y}}
  + \frac{-126 y^4 + 1614 y^3 - 926 y^2 + 14 y}{81(y - 1)^5} \ln y \right.
  \nnl[2mm] & \tSt \left.
  + \frac{1202 y^4 - 7569 y^3 + 5436 y^2 - 797 y}{243(y - 1)^4} 
  + \left[ \frac{12 y^4 + 92 y^3 - 56 y^2}{9(y - 1)^5} \ln y
         + \frac{28 y^4 - 270 y^3 + 36 y^2 + 62 y}{27(y - 1)^4} \right] L_t \right\},
\\[6mm]
  {[B_9^\lbl]}^1_H &= {[B_{10}^\lbl]}^1_H = 0,
\\[6mm]
  {[C_9^\lbl]}^1_H &= \tSt
   \frac{M_H^2}{8 M_W^2} \cot^2\beta \left\{ 
    \frac{-8 y^3 + 16 y^2}{(y - 1)^2} \Li{1 - \frac{1}{y}}
  + \frac{-24 y^3 + 88 y^2}{3(y - 1)^3} \ln y \right.
  + \frac{32 y^3 - 96 y^2}{3(y - 1)^2}
  \nnl[2mm] & \tSt \left.
  + \left[ \frac{16 y^2}{(y - 1)^3} \ln y
         + \frac{8 y^3 - 24 y^2}{(y - 1)^2} \right] L_t \right\}, 
\\[6mm] 
  {[D_9]}^1_H &= \tSt
  \cot^2\beta \left\{ 
    \frac{380 y^4 - 528 y^3 + 72 y^2 + 128 y}{81(y - 1)^4} \Li{1 - \frac{1}{y}}
  + \frac{596 y^4 - 672 y^3 + 64 y^2 + 204 y}{81(y - 1)^5} \ln y \right.
  \nnl[2mm] & \tSt
  + \frac{-6175 y^4 + 9138 y^3 - 3927 y^2 - 764y}{729(y - 1)^4}
  \nnl[2mm] & \tSt \left.
  + \left[ \frac{432 y^4 - 456 y^3 + 40 y^2 + 128 y}{81(y - 1)^5} \ln y
         + \frac{-352 y^4 - 972 y^3 + 1944 y^2 - 1052 y}{243(y - 1)^4} \right] L_t \right\}, 
\\[6mm]
  {[E_4]}^1_H &= \tSt
  \cot^2\beta \left\{ 
    \frac{515 y^4 - 906 y^3 + 99 y^2 + 182 y}{54(y - 1)^4} \Li{1-\frac{1}{y}}
  + \frac{1030 y^4 - 2763 y^3 - 15 y^2 + 980 y}{108(y - 1)^5} \ln y \right.
  \nnl[2mm] & \tSt
  + \frac{-29467 y^4 + 68142 y^3 - 6717 y^2 - 18134 y}{1944(y - 1)^4} 
  \nnl[2mm] & \tSt \left.
  + \left[ \frac{-375 y^3 - 95 y^2 + 182 y}{54(y - 1)^5} \ln y
         + \frac{133 y^4 - 108 y^3 + 4023 y^2 - 2320 y}{324(y - 1)^4} \right] L_t \right\},
\\[6mm]
  {[F_8]}^1_H &= \tSt  
    \frac{-17 y^3 + 25 y^2 - 36 y}{3(y - 1)^3} \Li{1 - \frac{1}{y}}
  + \frac{-34 y^3 + 7 y^2 - 165 y}{6(y - 1)^4} \ln y
  + \frac{29 y^3 - 44 y^2 + 143 y}{4(y - 1)^3} 
  \nnl[2mm] & \tSt
  + \left[ \frac{-34 y^2 - 38 y}{3(y - 1)^4} \ln y
         + \frac{7 y^3 - 16 y^2 + 81 y}{3(y - 1)^3} \right] L_t
  \nnl[2mm] & \tSt
 + \cot^2\beta \left\{
    \frac{-13 y^4 + 17 y^3 - 30 y^2}{3(y - 1)^4} \Li{1 - \frac{1}{y}}
  + \frac{-468 y^4 + 321 y^3 - 2155 y^2 - 2 y}{108(y - 1)^5} \ln y  \right.
  \nnl[2mm] & \tSt \left.
  + \frac{4451 y^4 - 7650 y^3 + 18153 y^2 - 1130 y}{648(y - 1)^4} 
  + \left[ \frac{-17 y^3 - 31 y^2}{3(y - 1)^5} \ln y 
         + \frac{7 y^4 - 18 y^3 + 261 y^2 + 38 y}{18(y - 1)^4} \right] L_t \right\},
\\[6mm]
  {[G_3]}^1_H &= \tSt
  \cot^2\beta \left\{ 
    \frac{10 y^4 + 30 y^2 - 20 y}{27(y - 1)^4} \Li{1-\frac{1}{y}}
  + \frac{30 y^3 - 66 y^2 - 56 y}{81(y - 1)^4} \ln y
  + \frac{6 y^3 - 187 y^2 + 213 y}{81(y - 1)^3} \right.
  \nnl[2mm] & \tSt \left.
  + \left[ \frac{-30 y^2 + 20 y}{27(y - 1)^4} \ln y
         + \frac{-35 y^3 + 145 y^2 - 80 y}{81(y - 1)^3} \right] L_t \right\}.
\end{align}
The following functions $[X]^1_H$ have been calculated previously:
$[A_7]^1_H$ and $[F_8]^1_H$ in \cite{CDGG:npb527,BG:prd58,BMU:npb567}
and $[B_{10}^\lbl]^1_H$ and $[C_9^\lbl]^1_H$ in \cite{BBKU:npb630}. The function
$[D_9]^1_H$ has been calculated in \cite{B:PhD} and confirmed in
\cite{SGST:hepph0407323}.  The results for the functions $[E_4]^1_H$
and $[G_3]^1_H$ are {\em new}. Note that $[B_9^\lbl]^1_H$ and
$[B_{10}^\lbl]^1_H$ vanish due to the approximation of vanishing lepton
masses.

%
%
%%%%%%%%%%%%%%%%%%%%%%%%%%%%%%%%%%%%%%%%%%%%%%
\subsection{\boldmath{$i=\tilde\chi$} -- ``chargino -- up squark''}

The evaluation of Feynman diagrams contributing to $b\to s+$(light particles)
Greens functions within the MSSM mediated by ``chargino -- up squark'' loops and
denoted by the index $i=\cIno$ in (\ref{Xfunctions}) yields
\begin{align}
  {[A_7]}^0_\cIno &= 
    \kappa \sum_{i=1}^2 \sum_{a=1}^6 \frac{M_W^2}{M_{\cIno_i}^2} \left\{ 
      \XX{i}{U_L}{2a}{\dagger} \XX{i}{U_L}{a3}{}\; h_1^{(0)}(y_{ai}) +
      \frac{M_{\cIno_i}}{m_b} \XX{i}{U_L}{2a}{\dagger} \XX{i}{U_R}{a3}{}\; h_2^{(0)}(y_{ai})
    \right\},     
\\[1mm]
  {[B_{9,10}^\lbl]}^0_\cIno &=  
   \mp \kappa \frac{M_W^2}{2 g_2^2} \sum_{i,j=1}^2 \sum_{a=1}^6 \sum_{b=1}^3
   \frac{ \XX{j}{U_L}{2a}{\dagger} \XX{i}{U_L}{a3}{} }{M^2_{\cIno_i}} \nnl
  & \times \left\{ \frac{1}{2} \XX{i}{N_L}{lb}{\dagger} \XX{j}{N_L}{bl}{} 
            f_5^{(0)}(x_{ji}, y_{ai}, v_{bi}) \mp 
            \XX{i}{N_R}{lb}{\dagger} \XX{j}{N_R}{bl}{}
           \sqrt{x_{ji}} f_6^{(0)}(x_{ji}, y_{ai}, v_{bi}) 
     \right\},
\\[1mm]
  {[C_9^\lbl]}^0_\cIno &=  
    {[C_L^\nbn]}^0_\cIno,
\\[6mm]
  {[D_9]}^0_\cIno &=  
    \kappa \sum_{i=1}^2 \sum_{a=1}^6 \frac{M_W^2}{M_{\cIno_i}^2} 
      \XX{i}{U_L}{2a}{\dagger} \XX{i}{U_L}{a3}{}\; h_3^{(0)}(y_{ai}), 
\\[5mm]
  {[E_4]}^0_\cIno &=
    \kappa \sum_{i=1}^2 \sum_{a=1}^6 \frac{M_W^2}{M_{\cIno_i}^2}
      \XX{i}{U_L}{2a}{\dagger} \XX{i}{U_L}{a3}{}\; h_4^{(0)}(y_{ai}), 
\\[5mm]
  {[F_8]}^0_\cIno &=
    \kappa \sum_{i=1}^2 \sum_{a=1}^6 \frac{M_W^2}{M_{\cIno_i}^2} \left\{ 
      \XX{i}{U_L}{2a}{\dagger} \XX{i}{U_L}{a3}{}\; h_5^{(0)}(y_{ai}) +
      \frac{M_{\cIno_i}}{m_b} \XX{i}{U_L}{2a}{\dagger} \XX{i}{U_R}{a3}{}\; h_6^{(0)}(y_{ai})
    \right\},     
\\
  {[A_7]}^1_\cIno &=
    \kappa \sum_{i=1}^2 \sum_{a=1}^6 \frac{M_W^2}{M_{\cIno_i}^2}\nnl[1mm]
  &\times\left\{ 
      \XX{i}{U_L}{2a}{\dagger} \XX{i}{U_L}{a3}{}\; h_1^{(1)}(y_{ai},L_{\uSq_a}) +
      \frac{M_{\cIno_i}}{m_b} \XX{i}{U_L}{2a}{\dagger} \XX{i}{U_R}{a3}{}\; h_2^{(1)}(y_{ai},L_{\uSq_a})
    \right\},
\\[5mm]
  {[B_{9,10}^\lbl]}^1_\cIno &=
   \mp \kappa \frac{M_W^2}{2 g_2^2} \sum_{i,j=1}^2 \sum_{a=1}^6 \sum_{b=1}^3
   \frac{ \XX{j}{U_L}{2a}{\dagger} \XX{i}{U_L}{a3}{} }{M^2_{\cIno_i}} \nnl
  & \times \left\{ \frac{1}{2} \XX{i}{N_L}{lb}{\dagger} \XX{j}{N_L}{bl}{} 
            \left[ f_8^{(1)}(x_{ji}, y_{ai}, v_{bi}) +
                   4 \left(1 + y_{ai} \frac{\partial}{\partial y_{ai}}\right) f_5^{(0)}(x_{ji}, y_{ai}, v_{bi})  
            \; L_{\uSq_a} \right] \right. \nnl
  & \left.\quad \mp 
            \XX{i}{N_R}{lb}{\dagger} \XX{j}{N_R}{bl}{} \sqrt{x_{ji}} 
            \left[ f_9^{(1)}(x_{ji}, y_{ai}, v_{bi}) +
                   4 \left(1 + y_{ai} \frac{\partial}{\partial y_{ai}}\right) f_6^{(0)}(x_{ji}, y_{ai}, v_{bi})
            \; L_{\uSq_a} \right] \right\}, \nnl
\\[1mm]
  {[C_9^\lbl]}^1_\cIno &= 
    {[C_L^\nbn]}^1_\cIno,
\\[5mm]
  {[D_9]}^1_\cIno &=
    \kappa \sum_{i=1}^2 \sum_{a=1}^6 \frac{M_W^2}{M_{\cIno_i}^2} 
      \XX{i}{U_L}{2a}{\dagger} \XX{i}{U_L}{a3}{}\; h_3^{(1)}(y_{ai},L_{\uSq_a}), 
\\[5mm]
  {[E_4]}^1_\cIno &=
    \kappa \sum_{i=1}^2 \sum_{a=1}^6 \frac{M_W^2}{M_{\cIno_i}^2} 
      \XX{i}{U_L}{2a}{\dagger} \XX{i}{U_L}{a3}{}\; h_4^{(1)}(y_{ai},L_{\uSq_a}), 
\\[5mm]
  {[F_8]}^1_\cIno &=
    \kappa \sum_{i=1}^2 \sum_{a=1}^6 \frac{M_W^2}{M_{\cIno_i}^2}\nnl[1mm]
 &\times \left\{ 
      \XX{i}{U_L}{2a}{\dagger} \XX{i}{U_L}{a3}{}\; h_5^{(1)}(y_{ai},L_{\uSq_a}) +
      \frac{M_{\cIno_i}}{m_b} \XX{i}{U_L}{2a}{\dagger} \XX{i}{U_R}{a3}{}\; h_6^{(1)}(y_{ai},L_{\uSq_a})
    \right\},
\\[5mm]
  {[G_3]}^1_\cIno &=
    \kappa \sum_{i=1}^2 \sum_{a=1}^6 \frac{M_W^2}{M_{\cIno_i}^2} 
      \XX{i}{U_L}{2a}{\dagger} \XX{i}{U_L}{a3}{}\; h_7^{(1)}(y_{ai},L_{\uSq_a}).
\end{align}
The following functions $[X]^1_\cIno$ have been calculated previously:
$[A_7]^1_\cIno$ and $[F_8]^1_\cIno$ in \cite{CDGG:npb534,BMU:npb567} and
$[B_{10}^\lbl]^1_\cIno$ and $[C_9^\lbl]^1_\cIno$ in \cite{BBKU:npb630}. The results for
the functions $[B_{9}^\lbl]^1_\cIno$, $[D_9]^1_\cIno,\, [E_4]^1_\cIno$ and $[G_3]^1_\cIno$ are
{\em new}. The expressions for the functions
${[C_L^\nbn]}^0_\cIno$ and ${[C_L^\nbn]}^1_\cIno$ correspond to the leading and
the next-to leading contributions to the function ${[C_L^\nbn]}_\cIno =
{[C_L^\nbn]}^0_\cIno + \alS/(4\pi) {[C_L^\nbn]}^1_\cIno$ given in (3.14) of
\cite{BBKU:npb630}.

%
%
%%%%%%%%%%%%%%%%%%%%%%%%%%%%%%%%%%%%%
\subsection{\boldmath{$i=4$} -- ``chargino -- up squark (quartic)''}

The evaluation of Feynman diagrams contributing to $b\to s+$(light particles)
Greens functions within the MSSM mediated by ``chargino -- up squark'' loops
containing the quartic squark vertex\footnote{ Strictly speaking these matching
contributions originate from the part of the quartic squark vertex proportional
to the strong coupling constant $\alS$.}  instead of gluon corrections and
denoted by the index $i=4$ in (\ref{Xfunctions}) yields
\begin{align}
  {[A_7]}^1_4 &=
    \kappa \sum_{i=1}^2 \sum_{a,b,c=1}^6 
    \frac{M_W^2}{M_{\cIno_i}^2}\; P^U_{ab}\, y_{bi}\, P^U_{bc}\; (1 + L_{\uSq_b}) 
   \nnl &
   \times \left\{ 
      \XX{i}{U_L}{2a}{\dagger} \XX{i}{U_L}{c3}{}\; 
         [ - q_1^{(1)}(y_{ai}, y_{ci}) + \frac{2}{3} q_2^{(1)}(y_{ai}, y_{ci})  ] \right.
   \nnl & \left. \quad
   +  \frac{M_{\cIno_i}}{m_b} \XX{i}{U_L}{2a}{\dagger} \XX{i}{U_R}{c3}{}\; 
         [ - q_3^{(1)}(y_{ai}, y_{ci}) + \frac{2}{3} q_4^{(1)}(y_{ai}, y_{ci})  ]
    \right\},
\\[5mm]
  {[B_{9,10}^\lbl]}^1_4 &=
    \pm \frac{\kappa}{2 g_2^2} \frac{4}{3} \sum_{i,j=1}^2 \sum_{f=1}^3 \sum_{a, b, c =1}^6
    \frac{M_W^2}{M_{\tilde\chi_i^{}}^2} \; P^U_{ab}\, y_{bi}\, P^U_{bc}\; (1+L_{\tilde u_b})\;
    \XX{j}{U_L}{2a}{\dagger} \XX{i}{U_L}{c3}{}
  \nnl
  & \times \left\{ \frac{1}{2} f^{(0)}_{9}(x_{ji}, y_{ai}, y_{ci}, v_{fi}) 
                    \XX{i}{N_L}{lf}{\dagger} \XX{j}{N_L}{fl}{} \right. \nnl
  & \left. \qquad \mp \sqrt{x_{ji}} f^{(0)}_{10}(x_{ji}, y_{ai}, y_{ci}, v_{fi})
    \XX{i}{N_R}{lf}{\dagger} \XX{j}{N_R}{fl}{} \right\},
\\[5mm]
  {[C_9^\lbl]}^1_4 &=
    \frac{\kappa}{6} \sum_{i,j=1}^2 \sum_{a, \dots, e,g,k=1}^6 
    P^U_{gk}\, y_{ki}\, P^U_{ke}\; (1+L_{\tilde u_k})\; \XX{j}{U_L}{2d}{\dagger} \XX{i}{U_L}{a3}{} 
  \nnl & \times \Bigg\{
    2\sqrt{x_{ji}} f_6^{(0)} (x_{ji}, y_{ai}, y_{di}) U_{j1}U_{i1}^\ast \delta_{ae} \delta_{gd} 
  \delta_{b1} \delta_{c1} 
  - f_5^{(0)} (x_{ji}, y_{ai}, y_{di}) V_{j1}^\ast V_{i1} \delta_{ae}\delta_{gd}
  \delta_{b1} \delta_{c1}
  \nnl &
  + f_5^{(0)} (y_{ai}, y_{bi}, y_{ci}) (\Gamma^{U_{L}} {\Gamma^{U_{L}\dagger}})_{c b} 
    \delta_{ij}\delta_{ae}\delta_{bg} \delta_{cd} 
  + f_5^{(0)} (y_{ai}, y_{ci}, y_{di}) (\Gamma^{U_{L}} {\Gamma^{U_{L}\dagger}})_{c b}
    \delta_{ij}\delta_{ab}\delta_{ce}\delta_{dg}\Bigg\},
\\[1mm]
  {[D_9]}^1_4 &=
    \kappa \sum_{i=1}^2 \sum_{a,b,c=1}^6 \frac{M_W^2}{M_{\cIno_i}^2}\; 
    P^U_{ab}\, y_{bi}\, P^U_{bc}\; (1 + L_{\uSq_b})\;
    \XX{i}{U_L}{2a}{\dagger} \XX{i}{U_L}{c3}{}\; q_5^{(1)}(y_{ai}, y_{ci}), 
\\[5mm]
  {[E_4]}^1_4 &=
    \kappa \sum_{i=1}^2 \sum_{a,b,c=1}^6 \frac{M_W^2}{M_{\cIno_i}^2}\; 
    P^U_{ab}\, y_{bi}\, P^U_{bc}\; (1 + L_{\uSq_b})\;
    \XX{i}{U_L}{2a}{\dagger} \XX{i}{U_L}{c3}{}\; q_6^{(1)}(y_{ai}, y_{ci}), 
\\[5mm]
  {[F_8]}^1_4 &=
    \kappa \sum_{i=1}^2 \sum_{a,b,c=1}^6 
    \frac{M_W^2}{M_{\cIno_i}^2}\; P^U_{ab} y_{bi}\, P^U_{bc}\; (1 + L_{\uSq_b})\; 
   \nnl  &
   \times \left\{ 
      \XX{i}{U_L}{2a}{\dagger} \XX{i}{U_L}{c3}{}\; q_2^{(1)}(y_{ai}, y_{ci}) +
      \frac{M_{\cIno_i}}{m_b} \XX{i}{U_L}{2a}{\dagger} \XX{i}{U_R}{c3}{}\; q_4^{(1)}(y_{ai}, y_{ci})
    \right\},
\\[5mm]
  {[G_3]}^1_4 &= 0.
\end{align}
The following functions $[X]^1_4$ have been calculated previously: $[A_7]^1_4$
and $[F_8]^1_4$ in \cite{BMU:npb567} and $[B_{10}^\lbl]^1_4$ and
$[C_9^\lbl]^1_4$ in \cite{BBKU:npb630}. The result for the functions
$[B_9^\lbl]^1_4$, $[D_9]^1_4,\, [E_4]^1_4$ and $[G_3]^1_4$ are {\em new}.

%%%%%%%%%  app. auxiliary functions  %%%%%%%%%%%%%%%%%%%%%%%%%%%%%%%%%%%%%%%%%%%
\section{\large Auxiliary functions}\label{aux:funcs}

Here we present explicit formulae for the loop functions $h_i^{(0)}(x)$,
$h_i^{(1)}(x)$ and $q_i^{(1)}(x,y)$ introduced in \ref{app:wil:coeff}. They read
\begin{align}
  h_1^{(0)}(x) &= \tSt
    \frac{3 x^2 - 2 x}{3 (x - 1)^4} \ln x + \frac{-8 x^2 - 5 x + 7}{18 (x - 1)^3},
\\[6mm]
  h_2^{(0)}(x) &= \tSt
    \frac{-6 x^2 + 4 x}{3 (x - 1)^3} \ln x + \frac{7 x - 5}{3 (x - 1)^2},
\\[6mm]
  h_3^{(0)}(x) &= \tSt
    \frac{-6 x^3 + 9 x^2 - 2}{9 (x - 1)^4} \ln x + \frac{52 x^2 - 101 x + 43}{54 (x - 1)^3},
\\[6mm]
  h_4^{(0)}(x) &= \tSt
    \frac{-1}{3 (x - 1)^4} \ln x + \frac{2 x^2 - 7 x + 11}{18 (x - 1)^3}, 
\\[6mm]
  h_5^{(0)}(x) &= \tSt
    \frac{-x}{(x - 1)^4} \ln x + \frac{-x^2 + 5 x + 2}{6 (x - 1)^3},
\\[6mm]
  h_6^{(0)}(x) &= \tSt
    \frac{2 x}{(x - 1)^3} \ln x + \frac{-x - 1}{(x - 1)^2},
\\[8mm]
  h_1^{(1)}(x,y) &= \tSt
    \frac{-48 x^3 - 104 x^2 + 64 x}{9 (x - 1)^4} \Li{1 - \frac{1}{x}}
  + \frac{-378 x^3 - 1566 x^2 + 850 x + 86}{81 (x - 1)^5} \ln x
  \nnl[2mm] & \tSt
  + \frac{2060 x^3 + 3798 x^2 - 2664 x - 170}{243 (x - 1)^4}
  + \left[ \frac{12 x^3 - 124 x^2 + 64 x}{9 (x - 1)^5} \ln x
         + \frac{-56 x^3 + 258 x^2 + 24 x - 82}{27 (x - 1)^4} \right] y,
\\[6mm]
  h_2^{(1)}(x,y) &= \tSt
    \frac{224 x^2 - 96 x}{9 (x - 1)^3} \Li{1 - \frac{1}{x}}
  + \frac{-24 x^3 + 352 x^2 - 128 x - 32}{9 (x - 1)^4} \ln x  
  + \frac{-340 x^2 + 132 x + 40}{9 (x - 1)^3} 
  \nnl[2mm] & \tSt
  + \left[ \frac{-24 x^3 + 176 x^2 - 80 x}{9 (x - 1)^4} \ln x
         + \frac{-28 x^2 - 108 x + 64}{9 (x - 1)^3} \right] y,
\\[6mm]
  h_3^{(1)}(x,y) &= \tSt
    \frac{32 x^3 + 120 x^2 - 384 x + 128}{81 (x - 1)^4} \Li{1 - \frac{1}{x}}
  + \frac{-108 x^4 + 1058 x^3 - 898 x^2 - 1098 x + 710}{81 (x - 1)^5} \ln x 
  \nnl[2mm] & \tSt
  + \frac{-304 x^3 - 13686 x^2 + 29076 x - 12062}{729 (x - 1)^4} 
  \nnl[2mm] & \tSt
  + \left[ \frac{540 x^3 - 972 x^2 + 232 x + 56}{81 (x - 1)^5} \ln x
         + \frac{-664 x^3 + 54 x^2 + 1944 x - 902}{243 (x - 1)^4} \right] y, 
\\[6mm]
  h_4^{(1)}(x,y) &= \tSt
    \frac{-562 x^3 + 1101 x^2 - 420 x + 101}{54 (x - 1)^4} \Li{1 - \frac{1}{x}}
  + \frac{-562 x^3 + 1604 x^2 - 799 x + 429}{54 (x - 1)^5} \ln x 
  \nnl[2mm] & \tSt
  + \frac{17470 x^3 - 47217 x^2 + 31098 x - 13447}{972 (x - 1)^4} 
  + \left[ \frac{89 x + 55}{27 (x - 1)^5} \ln x 
         + \frac{38 x^3 - 135 x^2 + 54 x - 821}{162 (x - 1)^4} \right] y, 
\\[6mm]
  h_5^{(1)}(x,y) &= \tSt
    \frac{9 x^3 + 46 x^2 + 49 x}{6 (x - 1)^4} \Li{1 - \frac{1}{x}}
  + \frac{81 x^3 + 594 x^2 + 1270 x + 71}{54 (x - 1)^5} \ln x 
  \nnl[2mm] & \tSt
  + \frac{-923 x^3 - 3042 x^2 - 6921 x - 1210}{324 (x - 1)^4} 
  + \left[ \frac{10 x^2 + 38 x}{3 (x - 1)^5} \ln x
         + \frac{-7 x^3 + 30 x^2 - 141 x - 26}{9 (x - 1)^4} \right] y, 
\\[6mm]
  h_6^{(1)}(x,y) &= \tSt
    \frac{-32 x^2 - 24 x}{3 (x - 1)^3} \Li{1 - \frac{1}{x}}
  + \frac{-52 x^2 - 109 x - 7}{3 (x - 1)^4} \ln x
  + \frac{95 x^2 + 180 x + 61}{6 (x - 1)^3} 
  \nnl[2mm] & \tSt
  + \left[ \frac{-20 x^2 - 52 x}{3 (x - 1)^4} \ln x 
         + \frac{-2 x^2 + 60 x + 14}{3 (x - 1)^3} \right] y, 
\\[6mm]
  h_7^{(1)}(x,y) &= \tSt
    \frac{-20 x^3 + 60 x^2 - 60 x - 20}{27 (x - 1)^4} \Li{1 - \frac{1}{x}}
  + \frac{-60 x^2 + 240 x + 4}{81 (x - 1)^4} \ln x
  + \frac{132 x^2 - 382 x + 186}{81 (x - 1)^3} 
  \nnl[2mm] & \tSt
  + \left[ \frac{20}{27 (x - 1)^4} \ln x
         + \frac{-20 x^2 + 70 x - 110}{81 (x - 1)^3} \right] y,
\\[8mm]
  q_1^{(1)}(x,y) &= \tSt
    \frac{4}{3 (x-y)} \left[ \frac{x^2 \ln x}{(x - 1)^4} 
                           - \frac{y^2 \ln y}{(y - 1)^4} \right]
  + \frac{4 x^2 y^2 + 10 x y^2 - 2 y^2 + 10 x^2 y - 44 x y + 10 y - 2 x^2 + 10 x + 4}
         {9 (x - 1)^3 (y - 1)^3},
\\[6mm] 
  q_2^{(1)}(x,y) &= \tSt
    \frac{4}{3 (x-y)} \left[ \frac{x \ln x}{(x - 1)^4} 
                           - \frac{y \ln y}{(y - 1)^4} \right]
  + \frac{-2 x^2 y^2 + 10 x y^2 + 4 y^2 + 10 x^2 y - 20 x y - 14 y + 4 x^2 - 14 x + 22}
         {9 (x - 1)^3 (y - 1)^3},
\\[6mm] 
  q_3^{(1)}(x,y) &= \tSt
    \frac{8}{3 (x-y)} \left[ \frac{-x^2 \ln x}{(x - 1)^3}
                           + \frac{ y^2 \ln y}{(y - 1)^3} \right]
  + \frac{-12 x y + 4 y + 4 x + 4}{3 (x - 1)^2 (y - 1)^2},
\\[6mm] 
  q_4^{(1)}(x,y) &= \tSt
    \frac{8}{3 (x-y)} \left[ \frac{-x \ln x}{(x - 1)^3}
                           + \frac{ y \ln y}{(y - 1)^3} \right]
  + \frac{-4 x y - 4 y - 4 x + 12}{3 (x - 1)^2 (y - 1)^2},
\\[6mm]
  q_5^{(1)}(x,y) &= \tSt
    \frac{4}{27 (x-y)} \left[ \frac{(6 x^3 - 9 x^2 + 2) \ln x}{(x - 1)^4} 
                            - \frac{(6 y^3 - 9 y^2 + 2) \ln y}{(y - 1)^4} \right]
  \nnl[2mm] & \tSt
  + \frac{104 x^2 y^2 - 202 x y^2 + 86 y^2 - 202 x^2 y + 380 x y - 154 y + 86 x^2 - 154 x + 56}
          {81 (x - 1)^3 (y - 1)^3},
\\[6mm]
  q_6^{(1)}(x,y) &= \tSt
    \frac{4}{9 (x-y)} \left[ \frac{\ln x}{(x - 1)^4} - \frac{\ln y}{(y - 1)^4} \right]
  + \frac{4 x^2 y^2 - 14 x y^2 + 22 y^2 - 14 x^2 y + 52 x y - 62 y + 22 x^2 - 62 x + 52}
         {27 (x - 1)^3 (y - 1)^3}.
\end{align}
The functions $f^{(0)}_5$, $f^{(0)}_6$, $f^{(0)}_9$, $f^{(0)}_{10}$, $f^{(1)}_8$
and $f^{(1)}_9$ can be found in Appendix B of \cite{BBKU:npb630}.

%%%%%%%%%%  referencies %%%%%%%%%%%%%%%%%%%%%%%%%%%%%%%%%%%%%%%%%%%%%%%%%%%%%%%

\setlength{\baselineskip}{2 mm}

\end{document}